\documentclass[11pt,a4paper]{article}
\pdfoutput=1 
\usepackage{booktabs}
\usepackage[english]{babel}
\usepackage[utf8]{inputenc}
\usepackage{amsmath,amssymb,amsbsy,amstext, amsthm} 
\usepackage{hyperref}
\usepackage[numbers,sort&compress]{natbib}
\usepackage{graphicx}
\usepackage{amsfonts}
\usepackage{amssymb}
\usepackage{upgreek}
\usepackage{exscale,relsize}
\usepackage[makeroom]{cancel}
\usepackage{soul}
\usepackage{bbold}
\usepackage{bbm}
\usepackage{color}
\usepackage[small]{caption}

\newcommand{\nn}{\nonumber}

\newcommand{\be}{\begin{equation}}
\newcommand{\ee}{\end{equation}}
\newcommand{\bea}{\begin{eqnarray}}
\newcommand{\eea}{\end{eqnarray}}
\newcommand{\bk}{\boldsymbol{k}}

\newcommand{\bq}{\boldsymbol{q}}
\newcommand{\bg}{\boldsymbol{g}}
\newcommand{\bx}{\boldsymbol{x}}
\newcommand{\by}{\boldsymbol{y}}

\newcommand{\bv}{\boldsymbol{v}}
\def\beq{\begin{equation}}
\def\eeq{\end{equation}}
\makeatletter
\newlength{\apb@width}
\newcommand{\autoparbox}[2][c]{\settowidth{\apb@width}{#2}\parbox[#1]{\apb@width}{#2}}

\makeatother

\numberwithin{equation}{section}

\begin{document}

\begin{titlepage}

\setcounter{page}{1} \baselineskip=15.5pt \thispagestyle{empty}
\begin{flushright}
CERN-PH-TH-2015-198\\
DESY 15-148\\
ICTP-SAIFR 15-127\\
\end{flushright}
\bigskip\

\begin{center}

{\fontsize{20}{28}\selectfont  \sffamily \bfseries 
On the Soft Limit of the \\  Large Scale Structure Power Spectrum:\\ \vskip 0.5 cm  UV Dependence} 

\end{center}

\vspace{0.2cm}

\begin{center}
{\fontsize{13}{30}\selectfont  Mathias Garny$^1$, Thomas~Konstandin$^2$, Rafael~A. Porto$^3$ and Laura~Sagunski$^2$}
\end{center}

\begin{center}
\textsl{$^1$ CERN Theory Division, CH-1211 Geneva 23, Switzerland}
\vskip7pt
\textsl{$^2$ Deutsches Elektronen-Synchrotron DESY, Theory Group, D-22603 Hamburg, Germany}
\vskip7pt
\textsl{$^3$ ICTP South American Institute for Fundamental Research, 01140-070 S\~ao Paulo, SP Brazil}\\
\end{center}

\vspace{1.2cm}
\hrule \vspace{0.3cm}
\noindent {\sffamily \bfseries Abstract} \\[0.04cm]
We derive a non-perturbative equation for the large scale structure power spectrum of long-wavelength modes. Thereby, we use an operator product expansion together with relations between the three-point function and power spectrum in the soft limit. The resulting equation encodes the coupling to ultraviolet (UV) modes in two time-dependent coefficients, which may be obtained from response functions to (anisotropic) parameters, such as spatial curvature, in a modified cosmology. We argue that both depend weakly on fluctuations deep in the UV. As a byproduct, this implies that the renormalized leading order coefficient(s) in the effective field theory (EFT) of large scale structures receive most of their contribution from modes close to the non-linear scale. Consequently, the UV dependence found in explicit computations within standard perturbation theory stems mostly from counter-term(s). We confront a simplified version of our non-perturbative equation against existent numerical simulations, and find good agreement within the expected uncertainties. Our approach can in principle be used to precisely infer the relevance of the leading order EFT coefficient(s) using small volume simulations in an `anisotropic separate universe' framework. Our results suggest that the importance of these coefficient(s) is a $ \sim 10 \%$ effect, and plausibly smaller. 
\vskip 10pt
\hrule

\bigskip

\end{titlepage}

\tableofcontents

\newpage 

%
\section{Introduction and Summary}
%

The measurement of anisotropies in the cosmic microwave background (CMB) by the Planck collaboration has provided invaluable information about the origin of the seed of structure \cite{Ade:2015xua}, plausibly created during an early phase of accelerated expansion \cite{Ade:2015lrj}. While this is a remarkable achievement, many questions still remain open regarding the nature of an inflationary cosmology. Perhaps the most outstanding one is whether a slowly-rolling, weakly coupled, fundamental scalar field played the role of the Higgs mechanism for the early universe; or if the universe chose a different path, such as dynamics at strong coupling or supersymmetry, e.g. \cite{Bmode1,Dansusy,Arkani-Hamed:2015bza}. These possibilities remain viable candidates to play a role in the early universe, and have not been considerably hindered by the Planck data \cite{Bmode1}. In particular, the existent bounds on (equilateral) non-Gaussianity \cite{Ade:2015ava} are still above some well-motivated thresholds \cite{Bmode1,Baumann:2015nta}.\vskip 4pt

Large-scale structure (LSS) surveys are due to become the next leading probe for precision cosmology, providing new information about the history of the universe through ambitious observational programs, currently under way to make very precise measurements. Therefore, the analytic control for the largest number of modes in the process of structure formation will allow us to place better constraints on cosmological parameters, and to further our understanding on the seed of structure. Unfortunately, quantitative predictions even for the simplest of the cosmologies is a daunting task, due to the non-linear nature of dark matter clustering. Moreover, even if one adopts a numerical approach, including baryonic matter is far from straightforward.\vskip 4pt Hence, motivated by the arrival of large amounts of data, a renewed interest on analytic methods in LSS has resurfaced. These can be broadly classified in two categories. On the one hand, studying the impact of long-wavelength (IR) fluctuations on modes around the scale of baryon acoustic oscillations, see e.g. \cite{Crocce:2005xz,Bernardeau:2011vy, Blas:2013bpa,Carrasco:2013sva,left,Senatore:2014via,Ben-Dayan:2014hsa,Baldauf:2015xfa} and references therein. On the other hand, considerable efforts have also been directed towards understanding the impact of short-wavelength (UV) perturbations. Their treatment is further complicated by the non-linear dynamics on short scales. 
The effective field theory (EFT) of LSS \cite{Baumann:2010tm,eft2,left} has emerged as a useful tool to parameterize the imprint of UV modes in long-distance observables, e.g. \cite{Carrasco:2013mua,Assassi:2015jqa, Baldauf:2015tla,Baldauf:2015zga,Baldauf:2015aha,Foreman:2015,Vlah:2015sea,McQuinn:2015tva}. (See also \cite{Manzotti:2014loa} for a somewhat different implementation.)  Motivated by various results from $N$-body simulations \cite{Pueblas:2008uv, Nishimichi:2014rra}, there have also been attempts to reorganize the perturbative expansion, e.g. \cite{Blas:2013aba, Blas:2015tla}.\vskip 4pt
In this paper we pursue another road, and study the soft (i.e. long-wavelength) limit of the power spectrum non-perturbatively. This allows us to scrutinize the actual impact of UV fluctuations on the power spectrum at large scales. As a byproduct, we assess the UV dependence of the leading order --renormalized-- coefficient(s) in the EFT, which capture the `finite size' effects on long-distance observables \cite{nrgr, Baumann:2010tm,eft2,left}.\vskip 4pt

The dynamics of large scale structures may be cast in a compact form using the density/velocity doublet $\psi_a \sim (\delta,\theta)$, with $\theta = -\nabla\cdot {\bv}$, which obeys the equation~\cite{Bernardeau:2001qr}
\beq
\partial_\eta \psi_a(\bq,\eta)+\Omega_{ac}(\eta) \psi_c(\bq,\eta) =  \gamma_{acd}\!\left(\bq/2-\bk,\bq/2+\bk\right)\psi_c(\bq/2-\bk,\eta)\psi_d(\bq/2+\bk,\eta)\label{flow}\,.
\eeq
Here $(\Omega,\gamma)$ are known matrices that depend on the dynamics of dark matter clustering. (Notice we use slightly different conventions for the vertices, see sec.~\ref{sec:softfluid} for more details. Everywhere we use that repeated indices/wavevectors must be summed/integrated over.) Within the EFT framework, extra terms must be added in order to account for smoothed regions. The relevance of these extra terms on long-distance observables can be read off from the mismatch between solutions to \eqref{flow} and data, and/or simulations.\footnote{In principle, the mismatch is also due to the vorticity, i.e. ${\boldsymbol{w}} =\nabla\times \bv$, which we ignored in \eqref{flow}. The latter may be neglected in the $q\to 0$ limit \cite{Bernardeau:2001qr}, but matters on short scales, $k \gg q$. However, these effects will not modify our main results. See sec. \ref{sec:disc}.}\vskip 4pt When the equations in \eqref{flow} are solved using standard perturbative techniques, the coefficients in the EFT approach also include counter-terms. This is necessary to cancel {\it would-be} divergences introducing e.g. a cutoff. (The so-called `loop expansion' produces ill-defined integrals for different initial conditions \cite{eft2,left}.) Even after the cutoff dependence is removed by the counter-terms, the EFT coefficients do not necessarily vanish. The left over (renormalized) contribution is physical, and accounts for the finite size corrections to \eqref{flow}. 
\vskip 4pt

To obtain the dynamics of the power spectrum, $P_{ab}(q,\eta)$, we multiply both sides of the equation by $\psi_c(\bq,\eta)$ and compute the statistical average~\cite{Pietroni:2008jx}. While appealing, the resulting equations are not very useful to solve for $P_{ab}(q,\eta)$ in closed analytic form, especially because the equation for the bispectrum, $B_{abc}$, on the right-hand side depends on the four-point function, and so on and so forth. In fact, an infinite hierarchy of equations between the $n$-point functions unfolds. It is then desirable to have a non-perturbative expression which would only depend on the power spectrum itself, and derivatives thereof. Even though we do not expect such relationship to hold in general, there are instances in which non-perturbative expressions may be found in specific limits, in particular when one of the wavenumbers is taken to be soft, i.e. $q \to 0$. In such case, we can use a relation for the density and velocity fields that resembles the operator product expansion (OPE) in quantum field theory. Keeping only lower order terms, 
\beq
\label{eq:ope}
\psi_a (\bq/2-\bk,\eta) \psi_b (\bq/2+\bk,\eta)  \xrightarrow{k \gg q}   \Big(f_{ab}(k,\mu,\eta) + g_{ab}(k,\mu,\eta)\frac{q}{k} \Big) \psi_L(\bq,\eta) +\cdots \,,
\eeq
where $\psi_L(\bq,\eta)$ is the linearized long-wavelength perturbation in the $q\to 0$ limit, and
\bea
f_{ab}(k,\mu,\eta) &=& f^{(0)}_{ab}(k,\eta) + f^{(2)}_{ab}(k,\eta) Q_2(\mu)\,, \label{eq:fope}\\
g_{ab}(k,\mu,\eta) &=& f^{(1)}_{ab}(k,\eta) Q_1(\mu)+ f^{(3)}_{ab}(k,\eta) Q_{3}(\mu)\,, \label{eq:gope}
\eea
with $\mu \equiv \bk\cdot \bq/(kq)$, and $Q_\ell (\mu)$ the Legendre polynomials. 
This relation reflects our prejudice that short scale physics is influenced by short modes in a very specific way. One incarnation of this principle are the relations between correlators in the soft limit, e.g.~\cite{Ben-Dayan:2014hsa}.
Strictly speaking, the expression in \eqref{eq:ope} is only accurate to the order we work here. A more detailed derivation is given in sec.~\ref{sec:PS}.\vskip 4pt

Inserting this expression in \eqref{flow}, multiplying by $\psi_c(\bq,\eta)$ and performing the statistical average we obtain 
\bea
\partial_\eta \, P_{ab}\!\left(q,\eta\right) &=& -\Omega_{ac} P_{cb}\!\left(q,\eta\right) - 
\Omega_{bc} P_{ac}\!\left(q,\eta\right) \nn \\ 
&& \hskip -3 cm
- \frac{q^2}{2} \, P_L(q,\eta) \, 
\Bigg\{ \left(\begin{array}{cc}
0 & 1 \\\
1 & 2 \\
\end{array}\right) C_{22}(\eta)
+
\left(\begin{array}{cc}
2 & 1 \\
1 & 0 \\
\end{array}\right)C_{12}(\eta) \Bigg\} \, ,
\label{a3}
\eea
up to terms which are higher order in $q$. 
Here $P_L(q,\eta)$ represents the linear approximation, and 
\bea
C_{22}(\eta) &=& C^{(0)}_{22}(\eta)\\
C_{12}(\eta) &=& C^{(0)}_{12}(\eta) + C^{(1)}_{12}(\eta) + C^{(2)}_{12}(\eta) \label{c12}
\eea
with
\bea
C^{(0)}_{22}(\eta) &=& 4\pi \int  dk~f^{(0)}_{22}(k,\eta)~, \\
C^{(0)}_{12}(\eta) &=& -\frac{4\pi}{3}\int dk~f^{(0)}_{12}(k,\eta)~, \\ 
C^{(1)}_{12}(\eta) &=& \frac{8\pi}{3} \int dk~f^{(1)}_{12}(k,\eta)~,\\ 
C^{(2)}_{12}(\eta) &=& \frac{16\pi}{15} \int dk~f^{(2)}_{12}(k,\eta)~. 
\eea
Notice that the expansion in small $q$ is only viable when the integrals in~
\eqref{a3} are dominated by hard momenta $k \gg q$. We will see this explicitly in sec.~\ref{sec:softfluid}.

\vskip 4pt

We arrive then to the desired equation, written solely in terms of $P_{ab}(k,\eta)$ and a set of coefficients which depend on fluctuations on short scales. However, in order to use this expression and match these coefficients with data or simulations, it is useful to relate $C_{22}(\eta)$ and $C_{12}(\eta)$ to the power spectrum of hard modes. This can be achieved using soft limit relations between $(n+1)$- and $n$-point correlation functions at {\it equal} times. For the case of spherically symmetric soft modes, these were discussed in \cite{Ben-Dayan:2014hsa}, see also \cite{Baldauf:2011bh,Sherwin:2012nh,Valageas:2013zda,Kehagias:2013paa}. For instance, for the density contrast one finds that the bispectrum obeys
\bea
B_{111} \!\left(\bq,-\bk-\bq,\bk,\eta \right)^{\rm av} 
\xrightarrow{q \to 0}
&P_L (q,\eta)& \Big[ \left(1 - \frac{1}{3}k~\frac{\partial}{\partial k} -\frac13 \frac{\partial}{\partial \eta} \right) P^{K=0}_{11} (k,\eta) \nn \\
&& + \frac{5}{3} \frac{\partial}{\partial \kappa}\left.  P^K_{11}(k,\eta)\right|_{K=0}\Big]  \, , \label{eq:soft0}
\eea
in an Einstein de Sitter (EdS) cosmology, and after angular averaging. Here $\kappa=K/(a^2H^2)$, with $K$ the curvature of a hypothetical background cosmology, while $P^K_{11}(k,\eta)$ is the full non-perturbative density power spectrum in the presence of spatial curvature. Similar expressions hold including the velocity field \cite{Ben-Dayan:2014hsa} (for which $P_L(q,\eta)$ remains unchanged by construction.) The~generalization for background cosmologies other than EdS is straightforward.  Hence, expressions like \eqref{eq:soft0} allow us to connect $C^{(0)}_{ab}(\eta)$ to the power spectrum on short scales, albeit in a locally curved universe. We obtain, for example, 
\beq
\label{c22}
C^{(0)}_{22}(\eta) =   -4 \sigma_{22}^2(\eta) - \frac{\partial}{\partial\eta} \sigma_{22}^2(\eta)
+ \left. 5\frac{\partial}{\partial \kappa}\left(\sigma^K_{22}\right)^2(\eta)\right|_{K=0}~,
\eeq
where\footnote{Notice that the expression for $\sigma_{ab}$ is dimensionful, since it represents the variance in the `displacements'. We could instead, following the Lagrangian-space EFT approach (LEFT) \cite{left}, introduce the parameters $\epsilon^{\psi<}_{ab}(q,\eta) \equiv q^2 \int_{0}^{q} \frac{d^3\bk}{k^2} P_{ab}(k,\eta)$ and $\epsilon^{\psi>}_{ab}(q,\eta) \equiv q^2 \int_q^\infty \frac{d^3\bk}{k^2} P_{ab}(k,\eta)$.  However, we find that keeping the $q^2$ makes the soft limit more explicit, and the distinction is unnecessary (between $\epsilon^{\psi<}$ and $\epsilon^{\psi>}$) when $q \to 0$.} 
\beq
\label{eq:sig}
\left(\sigma^K_{ab}\right)^2(\eta) \equiv  \frac{4\pi}{3} \int dk~P^K_{ab}(k,\eta)~.
\eeq

The angular averaged expression does not constrain the $\ell=\{1,2\}$ multipoles. In order to extract this information we need to include directional soft modes. As we shall discuss in the next section, these can be also absorbed into a cosmological background, this time with the addition of anisotropies. (See for instance \cite{Baldauf:2011bh} for the case of a plane wave perturbation.) The final expression for the bispectrum in the soft limit is not incredibly illuminating. 
Because of the anisotropies, one observes two different expansion rates,
parallel and perpendicular to the directional soft mode $\bq$. Moreover, there will be a curvature term, $K_\parallel$, which only enters in the parallel direction. An angular average reproduces the expression in \eqref{eq:soft0}. 

At the end of the day we will obtain an evolution equation for the power spectrum in the soft limit in terms of UV parameters which depend on variations of the spectrum of hard modes in a hypothetical anisotropic universe. The $C^{(\ell)}_{ab}(\eta)$ are then expressed in terms of the variance of the displacements in the new cosmology, which includes a series of geometrical parameters, 
\beq
\label{eq:sig2}
\left(\sigma^{A}_{ab}\right)^2(\eta_\perp,\eta_\parallel,K_\parallel,\cdots) \equiv  \frac{1}{3} \int \frac{d^3k}{k^2}~P^A_{ab}(k_\parallel,k_\perp,\eta_\perp,\eta_\parallel,K_\parallel, \cdots)~,
\eeq
(where we added the superscript `$A$' to emphasize these are defined for an anisotropic cosmology, and $k_\parallel,k_\perp$ denote the projections relative to $\bq$.)\vskip 4pt

It is then straightforward to show, from the above analysis, that only $\sigma^A_{22(12)}$ contribute to the $C^{(\ell)}_{22(12)}$ entering in the equation \eqref{a3}
for the power spectrum at large scales. We will argue here that these parameters depend weakly on hard modes beyond the non-linear scale. In turn, this implies that the hard modes which contribute the most to the power spectrum in the soft limit live near the non-linear scale.\vskip 4pt

\begin{figure}
 \centering
\includegraphics[width=0.55\textwidth]{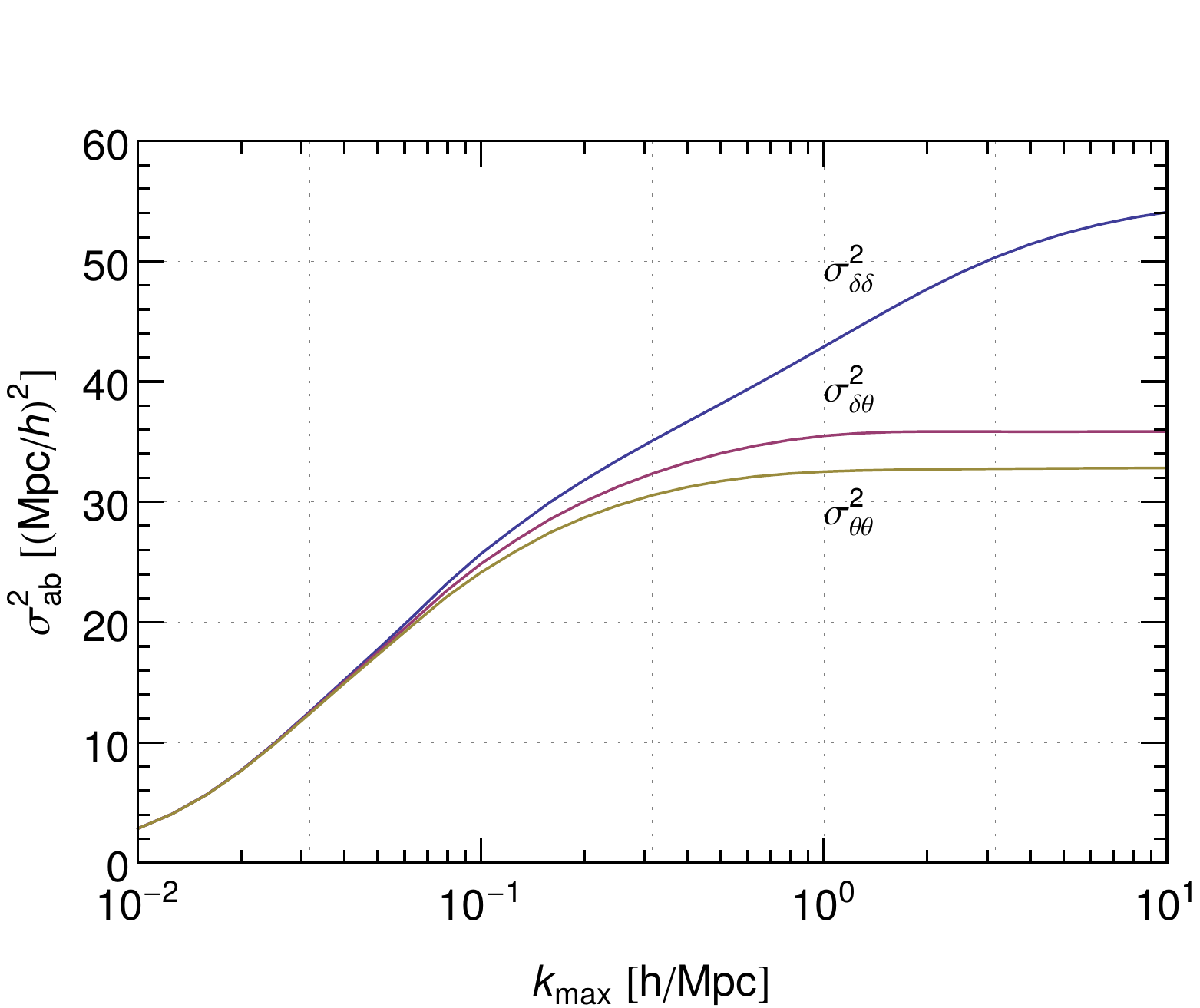}
\caption{Moments in \eqref{eq:sig3} as a function of $k_{max}$, at zero redshift. Results based on $N$-body simulations presented in \cite{Hahn:2014lca}. The dependence on short-distance scales is less relevant at earlier times. Note that $\sigma_{\delta\delta}\equiv \sigma_{11}, \sigma_{\delta\theta}\equiv \sigma_{12}, \sigma_{\theta\theta}\equiv \sigma_{22}$.}
\label{fig:sigma}
\end{figure}

Strong support for this claim follows from Fig.~\ref{fig:sigma}, where we plot $\sigma_{ab}^2(k_{\rm max})$ for our universe --without curvature-- as a function of a short-distance scale $k_{\rm max}$, defined as
\beq
\label{eq:sig3}
\sigma_{ab}^2(k_{\rm max}) \equiv \frac{4\pi}{3} \int_0^{k_{max}} dk~P_{ab}(k)~,
\eeq
at zero redshift.
We notice that $\sigma^2_{12}(k_{\rm max})$ and $\sigma^2_{22}(k_{\rm max})$ does not vary significantly for $k_{\rm max}$ greater than the non-linear scale $\Lambda_{\rm NL}~\simeq~0.5 \,h/$Mpc, in contrast to  $\sigma^2_{11}(k_{\rm max})$.\vskip 4pt 
 
While we do not expect the variance to be independent of the cosmology for $k_{\rm max} \ll \Lambda_{\rm NL}$, the curvature --and other geometrical anisotropic structure--  induced by the soft mode should not significantly affect $\sigma^A_{ab}$ on short scales. (These new parameters depend on the properties of a long-wavelength mode, $\delta_L(\bq) \ll 1$, with $|\bq| \ll \Lambda_{\rm NL}$.) 
In other words, we expect 
 \beq
\sigma^A_{ab}(k_{\rm max}) \simeq \sigma_{ab}(k_{\rm max})
 \eeq
for $k_{\rm max} \gg \Lambda_{\rm NL}$, with $\sigma^A_{ab}(k_{\rm max})$ defined similarly to \eqref{eq:sig3}. Then, the conclusion we draw from Fig.~\ref{fig:sigma} would also apply to the variance in a modified cosmology. Namely, the $\sigma_{22(12)}^A(k_{\rm max})$  display weak dependence on short-distance modes beyond the non-linear scale.\vskip 4pt

In order to translate this conclusion also to the UV parameters in \eqref{a3} one more step is required. That is because not only $\sigma^A_{22(12)}$ contribute, but also derivatives with respect to e.g. the curvature as in \eqref{c22}. However, for modes deep into the non-linear regime, we also expect the variance to be less sensitive to curvature and other geometrical parameters. For instance --for the case of a spherically symmetric soft mode-- one can show ($\kappa = K/(a^2H^2)$)
\beq 
\frac{\partial}{\partial \kappa} \left.P^K_{ab}(k,\eta)\right|_{K=0}  < P_{ab}(k,\eta)~,
\eeq
for $k \gg \Lambda_{\rm NL}$. This can be inferred from numerical simulations which test the so called {\it separate universe} approach \cite{Li:2014sga,Wagner:2014aka,Chiang:2014oga,Wagner:2015gva} (another name for absorbing a spherically symmetric long-wavelength perturbation into a locally curved background.) Since --on physical grounds-- we expect this behavior to be a general feature, also in the anisotropic case, we then conclude that both, $C_{22}(\eta)$ and $C_{12}(\eta)$, are weakly dependent on modes beyond the non-linear scale.  As a byproduct, this implies a  mild dependence on short modes beyond the non-linear scale for the renormalized coefficient(s) of the EFT approach at leading order in derivatives \cite{Baumann:2010tm,eft2,left}, as we shall discuss.\vskip 4pt 

The conclusion that the renormalized EFT coefficient(s) at leading order in the soft limit do not depend strongly on the UV modes does not necessarily mean they have to be small. In order to test the validity of \eqref{flow} as a correct description of the dynamics, hence assessing the size of the renormalized EFT corrections, we will perform a truncation of \eqref{a3} which is amenable for comparison with existent numerical simulations. 
Our truncation consists on keeping only the $C_{22}(\eta)$ contribution to \eqref{a3}. As we discuss, the rationale for neglecting $C_{12}(\eta)$ is motivated by the fact that it arises from higher multipoles (in a suitable basis), and its contribution is suppressed by a relative factor $\sim 20$ in perturbation theory. Moreover, we will be using the approximation (see \cite{Ben-Dayan:2014hsa} and \cite{Valageas:2013zda,Kehagias:2013paa} for details)
\beq
\label{disc1}
 \left. \frac{\partial}{\partial \kappa} \psi^K_a(k,\eta) \right|_{K=0} \simeq \frac{4}{7}
\left(\begin{array}{cc}
\partial_{\eta} & 0\\
0 & \,\partial_{\eta}+1
\end{array}\right)_{ab} \psi^K_b(k,\eta). 
\eeq
We do not expect this approximation to hold deep into the non-linear scale. As we argued, the curvature dependence should be less important the shorter the scales.  However, given the information in Fig.~\ref{fig:sigma}, the approximation in \eqref{disc1} may be in principle more reliable, since the relevant modes involved in \eqref{a3} do not live too far from the non-linear scale. This enables us to use \eqref{disc1}, and replace the derivative with respect to spatial curvature in \eqref{c22} by a time derivative, obtaining
\beq
\label{c22new}
C_{22}(\eta) = \frac{1}{7} \left( 12~\sigma^2_{22}(\eta) + 13 \frac{\partial}{\partial \eta}  \sigma^2_{22}(\eta) \right),
\eeq
in an EdS background cosmology, now with $K=0$.\vskip 4pt The resulting (simplified) equation allows us then to make a direct comparison with traditional numerical methods with relatively accurate success. We show that the solution to this truncated equation reproduces  numerical results within $\sim 10 \%$ accuracy at zero redshift, which is also the expected error due to the approximations. 
The matching becomes more accurate --percent-level-- at higher redshifts. Our analysis does not take into account the plausible scale-dependence  (through a {\it renormalization group flow}) of the renormalized EFT parameters. However, our results suggest that the importance of the leading renormalized coefficient(s) is at most a $\sim 10 \%$ effect, and plausibly smaller. \vskip 4pt

In the next sections we derive all these expressions in more detail. We first derive the coordinate transformation to a locally curved anisotropic background in sec. \ref{sec:direct}. In sec.~\ref{sec:PS}, we use this transformation to relate the soft limit of the bispectrum with the variance on short-scales. Later on we use these relations to extract the coefficients in the OPE expansion in terms of integrals of the variance in a modified cosmology. We then perform a truncation of \eqref{a3}, keeping only the $C_{22}(\eta)$ contribution and using the approximation in \eqref{disc1}. We solve this simplified equation and show that it fares relatively well against numerical simulations in sec. \ref{sec:num}. We also discuss the errors induced by applying \eqref{disc1} and ignoring $C_{12}(\eta)$. We conclude in sec.~\ref{sec:disc} with a discussion on the implications of our findings, and in particular for the EFT of LSS. Appendix \ref{app:C12} contains some numerical estimates for the impact of $C_{12}(\eta)$.

%
\section{Directional soft modes: A locally curved anisotropic universe}\label{sec:direct} 
%

In this section we display the coordinate transformation needed to absorb a directional soft mode into an anisotropic background. We will study this map up to order $(\bq\cdot \bx)^3$ in the soft mode, which is necessary to extract the $C_{12}(\eta)$ coefficient. We also discuss the impact in the soft limit relations between the bispectrum and power spectrum in a modified cosmology. Here we follow similar steps as in \cite{Ben-Dayan:2014hsa}, but extend the analysis to include a directional mode. We first study the mapping to a locally curved anisotropic universe and then apply this transformation to the soft limit of the bispectrum.  
(See also \cite{Dai:2015jaa} for a somewhat related discussion within the separate universe approach.)

\subsection{Newtonian mapping}

We start from a FRW cosmology in global coordinates that includes a long-wavelength density perturbation with wavevector $\bq$, 
\beq 
\delta_L(\bx,t)\simeq \frac12 \delta_L(t) e^{i \, \bq\cdot \bx} + \textrm{c.c.}~,
\eeq
such that 
\be
\Phi(\bx,t) \simeq \frac32 \frac{H^2 a^2}{q^2} 
\left[
\frac12 \delta_L(t) e^{i \, \bq\cdot \bx} + \textrm{c.c.}
\right] \, .
\ee

In the Newtonian approximation, we have
\beq
ds^2=-\bigl[ 1+ 2\Phi_L(\bx,t)\,\bigr] dt^2 + a^2(t)\bigl[1 - 2\Phi_L(\bx,t)\bigr] d\bx^2\,.
\label{eq:metricN}
\eeq 

We will search for a coordinate transformation of the form 
\bea
t &=& t_A + f(t_A, \bx_A) \, , \nn \\
\bx &=& \bx_A + \bg(t_A, \bx_A) \, ,
\label{eq:trafo} 
\eea
with coordinate dependent functions $f$ and $\bg$, such that one can absorb the long-wavelength perturbation into a locally curved Bondi-type metric in a series expansion in $\bq$. We added the subscript `$A$' to emphasize this corresponds to an anisotropic cosmology. We start with the first non-trivial order.

\subsubsection{Parallel curvature $\&$ scale factor}

After the transformation in \eqref{eq:trafo} is performed, we expect the metric to take the form
\begin{align}
\label{ds2}
ds^2 \equiv &- d t_A^2 
+ a_{\parallel}^2(t_A) d\bx_{A,\parallel}^2
+ a_{\perp}^2(t_A) d\bx_{A,\perp}^2 \nn \\
&- \frac12 \left(K_\parallel  \bx_{A,\parallel}^2(t_A)  + K_\perp(t_A)  \bx_{A,\perp}^2  \right) a^2(t_A) d\bx_A^2~,
\end{align}
plus ${\cal O}(\bx_A^3)$ corrections we will discuss momentarily. Notice that, at leading order in $\delta_L$, the distinction between $a_{\perp}$ and $a_{\parallel}$ is irrelevant in the term involving curvature. For notational purposes, we drop the subscript `$A$' in what follows.\vskip 4pt

The extra time-dependent functions, $(a_{\perp}, a_{\parallel}, K_{\perp}, K_{\parallel})$, are the scale factors and curvature parameters respectively. These are defined in the perpendicular and parallel directions with respect to the soft wavevector $\bq$, 
\bea
\bx_\parallel &=& \frac{\bx \cdot \bq}{q^2}~\bq~,\\ 
\bx_\perp &=& \bx - \bx_\parallel~.
\eea

The form of the time-time component of the metric enforces
\be
\dot f = - \Phi_L \, , 
\ee
while the vanishing time-space part yields
\be
a^2 \, \dot  \bg = \nabla f \, .
\ee
Hence we find 
\be
\label{eq:constr1}
f = - \int \frac{da}{a} \frac1H \Phi_L~,
\ee
and 
\be
\label{eq:constr2}
\bg = \int \frac{da}{a} \frac{1}{a^2 H} \, \nabla f \, + \boldsymbol{c} \,  .
\ee
We introduced a time-independent constant of integration for $\bg$, but this is not required for $f$. In EdS, where $\Phi_L \propto a^0$, $\delta_L \propto a$ and $H \propto a^{-3/2}$ we have
\be
f = -  \frac23 \frac1H \Phi_L~,
\ee
and 
\be
\bg  = -\frac23 \frac{\nabla \Phi_L}{a^2 H^2} + \boldsymbol{c} \, ,
\ee
and we consider only this simple case in the following.\vskip 4pt

Applying now the transformation (\ref{eq:trafo}) in the metric (\ref{eq:metricN})
one finds, in combination with the constraints (\ref{eq:constr1}) and (\ref{eq:constr2}),
\bea
ds^2 &\simeq& a^2 \, d\bx^2 \, \left(1 - 2 \Phi_L - 2 H \int \frac{da}{a} \frac1H \Phi_L \right) \nn \\
&&  +  a^2 \,  dx_i dx_j (\nabla_i g_j + \nabla_j g_i) \, .
\eea
Notice that the term
\be
2 \Phi_L + 2 H \int \frac{da}{a} \frac1H \Phi_L \simeq \frac{10}{3} \Phi_L,
\ee
is constant in time. Next we expand $\Phi$ in $\bq \cdot \bx$, 
\be
\Phi(\bx,t) \simeq \frac32 \frac{H^2 a^2}{q^2} 
\left[
\frac12 \delta_L(t) 
\left(1 + i \, \bq\cdot \bx - \frac12(\bq\cdot \bx)^2 - \frac{i}6 (\bq\cdot \bx)^3 + \cdots \right)
+ \textrm{c.c.}
\right]  \, .
\ee
Likewise the integration constant $\boldsymbol{c}$ can be expanded as 
\be
\label{boldc}
\boldsymbol{c} = 
c_1 \, \bx + c_2 \,  (\bx \cdot \bq) \bx+ c_3\bx^2 \bq+ {\cal O}(\bx^3) \, ,
\ee
which provides an additional contribution to the metric of the form
\bea
dx_i dx_j (\nabla_i c_j + \nabla_j c_i)  &=&  d\bx^2 (c_1 + c_2\, \bx \cdot \bq) \nn \\
&& +  \, dx_i dx_j x_i q_j (c_2 + 2 c_3) \, . 
\eea
We notice that the constant and linear term in $ds^2$ can be removed using \eqref{boldc}. 
The last remaining contribution comes from the time-dependent part of $\bg$, which leads to 
\be
\label{dxidxj}
dx_i dx_j (\nabla_i g_j + \nabla_j g_i) \ni 2 \frac{(\bq \cdot d\bx)^2}{q^2} \delta_L(t) \, .
\ee
Garnering the pieces together, we have succeeded in transforming the metric in \eqref{eq:metricN} into (restoring the labels)
\begin{align}
\label{ds2n}
ds^2 = - d t_A^2 
+ a^2 d \bx_A^2+ 2a^2\delta_L(t) \frac{(\bq \cdot d\bx_A)^2}{q^2} - \frac52  H^2 a^2 \delta_L (t) \frac{(\bq \cdot \bx_A)^2}{q^2} d\bx_A^2~,
\end{align}
which has the form in \eqref{ds2}, after the identification
\bea
\label{eq1a}
a_{\perp} &=& a \, ,  \\
K_\perp&=&0~,\\
a_{\parallel} &=& a \, \Big(1 - \delta_L(t)\Big) \, , \\
K_\parallel  &=& 5 H^2 a^2 \delta_L (t)  \, .
\label{eq1b}
\eea
The angular averaging for these expression is slightly subtle. The reason is because the above quantities are defined with respect to the soft wavevector, $\bq$, which would then be averaged over angles. It is clear, nonetheless, that after averaging  ($q^iq^j/q^2 \to \tfrac{1}{3} \delta^{ij}$), the expression in \eqref{ds2n} leads to the usual result \cite{Baldauf:2011bh,Ben-Dayan:2014hsa}
\be
K= \frac53 H^2 a^2 \delta_L  \, ,
\ee
and
\be
a_K=  a \, \left(1 - \frac13 \delta_L \right) \, ,
\ee
as expected.

\subsubsection{More anisotropy} \label{sec:direct2}

We now include an extra order in $q$, which is required to express the $C_{12}(\eta)$ coefficient in terms of derivatives of geometrical properties of the power spectrum in a modified cosmology. Notice that at this order anisotropies become important, and the local cosmology is not simply described in terms of tidal forces in an isotropic FRW universe.\vskip 4pt Like before, the integration constant $\boldsymbol{c}$ can be used to eliminate certain terms. For instance, it can be used to transform terms of the form $d\bx^2 \bx^2$ into $(d\bx \cdot \bx)^2$, etc. After the dust settles, we find the following expression for the metric
\begin{align}
\label{ds2n2}
ds^2 = &- d t_A^2 
+ a^2 d \bx_A^2+ 2a^2\delta_L(t) \frac{(\bq \cdot d\bx_A)^2}{q^2} - \frac52  H^2 a^2 \delta_L (t) \frac{(\bq \cdot \bx_A)^2}{q^2} d\bx_A^2~\nn\\
&- \frac53  H^2 a^2 \delta_L (t) \frac{(\bq \cdot \bx_A)^3}{q^2} d\bx_A^2 + 2 a^2 \delta_L(t)\frac{(\bq \cdot d\bx_A)^2}{q^2} (\bq \cdot \bx_A),
\end{align}
with the addition of two new terms beyond \eqref{ds2n}. This can be mapped into a metric of the form
\begin{align}
\label{ds2n3}
ds^2 \equiv &- d t_A^2 +
a^2\, d\bx_{A,\perp}^2 + \left(1+ b_\parallel \, q x_{A,\parallel} \right) a_{\parallel}^2 \, d\bx_{A,\parallel}^2
  \nn \\
&- \frac12 \left(K_\parallel + {\cal K}_\parallel \, q x_{A,\parallel}  \right) a^2 \, \bx_{A,\parallel}^2 \, d\bx_{K}^2~,
\end{align}
with the identification in \eqref{eq1a}-\eqref{eq1b}, together with
\bea
b_\parallel &=& 2 \delta_L(t)~, \\
{\cal K}_\parallel &=& \frac{10}{3} H^2 a^2 \delta_L(t)~.
\eea
Notice $b_\parallel a^2_\parallel \to b_\parallel a^2$, at leading order in $\delta_L$. These extra terms do not survive after angular averaging. As we shall argue, fortunately contributions involving $C_{12}$ are subleading, such that we may concentrate on the isotropic terms.  

\subsection{The soft limit of the bispectrum}

In the anisotropic background, the fluctuations $\psi^A_a(\bk,\eta)$ depend upon two scale factors and the curvature parameters $K_\parallel, {\cal K}_\parallel$. Also the power spectrum, $P_{ab}^A(\cdots)$, depends not only on the absolute value of the wavevector $\bk$ but on $\bk_\perp$ and $\bk_\parallel$ separately. Following the procedure outlined in \cite{Ben-Dayan:2014hsa}, we have for the soft limit of the bispectrum 
\bea
B_{111} \!\left(\bq,-\bk-\bq,\bk,\eta \right) \xrightarrow{q \to 0} 
&P_L (q,\eta)&
\Big[
\left(1 - \bk_\parallel \frac{\partial}{\partial \bk_\parallel}
 - \frac{\partial }{\partial \eta_\parallel}\right)P^A_{11}(\bk_\perp,\bk_\parallel,\eta_\perp,\eta_\parallel, K_\parallel)  \nn \\
&&  + 5 \frac{\partial}{\partial \kappa_\parallel} P^A_{11}(\bk_\perp,\bk_\parallel,\eta_\perp,\eta_\parallel, K_\parallel)
\Big]  +\cdots \, . \label{eq:soft32}
\eea
with $\eta_\parallel \equiv \log a_\parallel$ and $\kappa_\parallel = K_\parallel/(H^2a^2)$. The right side is understood to be evaluated at $\eta_\parallel = \eta_\perp = \eta$, $\bk_\parallel = \bk_\perp = \bk$ and 
$\kappa_\parallel = 0$ after the derivatives have been performed. The ellipses include higher order terms in $q$, which incorporate derivatives with respect to e.g. ${\cal K}_\perp$ in \eqref{ds2n2}. These are in principle required to extract the coefficients entering in \eqref{c12} for $C_{12}(\eta)$. To avoid a --rather cumbersome-- expression we do not write them down explicitly, but they can be straightforwardly obtained from \eqref{ds2n3}.\vskip 4pt On the other hand, only the $\ell =0$ contribution is needed to extract $C_{22}(\eta)$. The latter can be obtained from \eqref{eq:soft0}, which follows by averaging \eqref{eq:soft32} over angles, i.e.  \cite{Ben-Dayan:2014hsa}
\bea
B_{111} \!\left(\bq,-\bk-\bq,\bk,\eta \right)^{\rm av} 
\xrightarrow{q \to 0}
&P_L (q,\eta)& \Big[ \left(1 - \frac{1}{3}k~\frac{\partial}{\partial k} -\frac13 \frac{\partial}{\partial \eta} \right) P^{K=0}_{11} (k,\eta) \nn \\
&& + \frac{5}{3} \frac{\partial}{\partial \kappa}\left.  P^K_{11}(k,\eta)\right|_{K=0}\Big]  \, . \label{eq:soft}
\eea

\section{A non-perturbative equation for the power spectrum\label{sec:PS}}

In here we derive all the necessary ingredients that enter in the non-perturbative equation \eqref{a3} for the power spectrum at large scales. An intuitive way to derive this equation is to use an OPE for two fluctuations at nearby points. This will be done in sec.~\ref{sec:ope}. An explicit example of a product expansion is given by the non-perturbative relations between correlator functions in the soft limit, e.g. \cite{Ben-Dayan:2014hsa}. These relations can be used to fix the coefficients of the OPE in terms of the power spectrum on short scales, albeit in a modified cosmology. This will be discussed in sec.~\ref{sec:softfluid}. 

\subsection{Operator product expansion}\label{sec:ope}

For the theory of structure formation we may think of the gravitational potential, $\Phi(\bx,\eta)$, as the relevant degree of freedom from which a solution for $\psi_a(\bx,\eta)$ follows, e.g. by deriving the displacement fields that follow from the induced forces, i.e. $-\nabla \Phi(\bx,\eta)$, as a function of time. An OPE is then an expansion for product of $\psi_a$ fields at nearby points, 
\beq
\psi_a(\bx,\eta)\psi_b(\by,\eta) \xrightarrow{\bx \to \by} \sum_{\cal O} f^{\cal O}_{ab}(|\bx-\by|,\eta)~{\cal O}[\Phi,\partial\Phi,\cdots] \Big(\tfrac{1}{2}(\bx+\by),\eta\Big),
\eeq
as a function of (composite) {\it operators} that are built in terms of $\Phi(\bx,\eta)$ and its derivatives. 
In Fourier space it reads
\beq
\label{eq:ope2}
\psi_a (\bq/2-\bk,\eta) \psi_b (\bq/2+\bk,\eta)  \xrightarrow{k \gg q}  \sum_{\cal O} f^{\cal O}_{ab}(k,\eta)~{\cal O}(\bq,\eta).
\eeq
\vskip 4pt
For our purposes it turns out to be sufficient to keep only terms linear in $\Phi$, because of the smallness of the density perturbations (and gradients) in the $q \to 0$ limit. However, let us briefly mention some caveats of the OPE expansion in our setting. Fist of all, 
only a handful of examples are known to obey an OPE beyond perturbation theory. Moreover, for the theory of structure formation, an extra complication arises. This is due to the statistical properties of the initial state. In other words, the existence of different possible realization introduces {\it stochastic} terms which are not necessarily proportional to products of long-wavelength fields.\footnote{These are `contact-terms' in the jargon of quantum field theory.} However, for our purposes, we may ignore these terms, which are known to enter at higher orders in the evolution equation for the power spectrum, i.e. $q^4$ \cite{Peebles:1980}. (The relevance of these terms, as one approaches the non-linear scale, has been recently emphasized in \cite{Baldauf:2015tla,Baldauf:2015zga}.) Furthermore, the operators in the right-hand side of \eqref{eq:ope2} may themselves involve products of fields at the same point, which in turn need to be regularized/renormalized \cite{left}. However, the net effect for these extra terms is sub-leading, since they are suppressed by extra powers of $P_L$.
\vskip 4pt

Due to the equivalence principle, we know the expansion starts with --at least-- two derivatives of the gravitational potential. Thus, working in Fourier space and using isotropy, parity, and the equivalence principle, we need
\bea
{\cal O}_0 (\bq,\eta) &=& q^2 \Phi (\bq,\eta) \propto \delta(\bq,\eta)~, \\
{\cal O}_i (\bq,\eta) &=& q^2 q^i \Phi (\bq,\eta) \propto q^i \delta(\bq,\eta)~, \\
{\cal O}^{\rm TF}_{ij} (\bq,\eta) &=& (q^iq^j)_{\rm TF}~\Phi(\bq,\eta)~,\\ 
{\cal O}^{\rm TF}_{ijl} (\bq,\eta) &=& (q^i q^j q^l)_{\rm TF}~\Phi(\bq,\eta)~,
\eea
where TF stands for trace-free. (The traces renormalize the coefficients of tensor operators with fewer indices, and we do not include them for simplicity of notation.) Moreover, for the functions of the hard modes we can, in addition to $f^{{\cal O}_0}_{ab}(k,\eta)$, use a similar splitting into scalar functions:
\bea
f^{{\cal O}_i}_{ab;i}(k,\eta) &=& \frac{k^i}{k} f^{{\cal O}_i}_{ab}(k,\eta)~,\\
f^{{\cal O}^{\rm TF}_{ij}}_{ab;ij}(k,\eta) &=& \frac{(k^ik^j)_{\rm TF}}{k^2}~ f^{{\cal O}^{\rm TF}_{ij}}_{ab}(k,\eta)~,\\
f^{{\cal O}^{\rm TF}_{ijl}}_{ab;ijl}(k,\eta) &=& \frac{(k^i k^j k^l)_{\rm TF}}{k^3} f^{{\cal O}^{\rm TF}_{ijl}}_{ab}(k,\eta)~.
\eea
When the dust settles, after performing all the contractions, we wound up with
\beq
\label{EqOPE}
\psi_a (\bq/2-\bk,\eta) \psi_b (\bq/2+\bk,\eta)  \xrightarrow{k \gg q} \Big(f_{ab}(k,\mu,\eta) + g_{ab}(k,\mu,\eta)\frac{q}{k}\Big) \psi_L(\bq,\eta) +\cdots~,
\eeq
where $f_{ab}(k,\mu,\eta)$ and $g_{ab}(k,\mu,\eta)$ are polynomials in $\mu$, up to order $\mu^3$, with coefficients given by a combination of the scalar functions defined above. Higher orders in $\mu$ arise when expanding in higher orders in $q$. 
These functions can be re-written in terms of Legendre polynomials as in \eqref{eq:fope}-\eqref{eq:gope}:
\bea
f_{ab}(k,\mu,\eta) &=& f^{(0)}_{ab}(k,\eta) + f^{(2)}_{ab}(k,\eta) Q_{2}(\mu)~,\\
g_{ab}(k,\mu,\eta) &=& f^{(1)}_{ab}(k,\eta) Q_1(\mu)+ f^{(3)}_{ab}(k,\eta) Q_{3}(\mu)~.
\eea
The expression in \eqref{EqOPE} implies,
\beq
\label{EqOPE2}
B_{1ab}(-\bq, \bq/2-\bk,\bq/2+\bk) \xrightarrow{q \to 0} \Big(f_{ab}(k,\mu,\eta) + g_{ab}(k,\mu,\eta)\frac{q}{k} \Big) P_L(q,\eta) + \cdots ~,
\eeq
for the bispectrum in the soft limit, and also with $B_{1ab} \to B_{2ab}$. (Recall $\psi_a$ is defined such that $P_{11}(q,\eta) = P_{22}(q,\eta) =P_L(q,\eta)$ at linear order \cite{Pietroni:2008jx}.) Notice, as expected, we do not have a term proportional to $1/q$. This is related to the so-called {\it consistency condition} for large scale structure correlation functions, which vanishes when evaluated at equal times. This is nothing but a reflection of the equivalence principle, see~\cite{Creminelli:2013nua} and references therein. 

\subsection{The fluid equations in the soft limit}\label{sec:softfluid}

The dynamics in \eqref{flow} leads to
\bea
\label{eq:P1N}
\partial_\eta \, P_{ab}\!\left(q,\eta\right) =
&-&\Omega_{ac} P_{cb}\!\left(q,\eta\right)
- \Omega_{bc} P_{ac}\!\left(q,\eta\right)  \\
   &+& 
\gamma_{bcd}\!\left(-\bq/2-\boldsymbol{k},
\boldsymbol{k}+\boldsymbol{q}/2\right)B_{acd}\!\left(\boldsymbol{q},-\bq/2-\boldsymbol{k
},\boldsymbol{k}-\boldsymbol{q}/2,\eta\right) \nn \\
&+& \gamma_{acd}\!\left(\bq/2-\boldsymbol{k},\boldsymbol{k}
+\boldsymbol{q}/2\right)B_{bdc}\!\left(-\boldsymbol{q},\bq/2-\boldsymbol{k},\bk+\boldsymbol{q}/2,\eta\right)\,,
\nn
\eea
for the evolution equation for the power spectrum \cite{Bernardeau:2001qr}. Moreover,
\begin{equation}
\begin{alignedat}{1}\gamma_{121}\!\left(\boldsymbol{k},
\boldsymbol{p}\right)=\gamma_{112}\!\left(\boldsymbol{p},
\boldsymbol{k}\right) &
=\,\alpha\!\left(\boldsymbol{k},\boldsymbol{p}\right)\,,\\
\gamma_{222}\!\left(\boldsymbol{k},\boldsymbol{p}\right) &
=\,\beta\!\left(\boldsymbol{k},\boldsymbol{p}\right)\,,
\end{alignedat}
\label{eq: gamma elements}
\end{equation}
with
\begin{equation}
\alpha\!\left(\boldsymbol{k},\boldsymbol{p}\right)=\frac{\left(\boldsymbol{
k}+\boldsymbol{p}\right)\cdot\boldsymbol{k}}{2k^{2}}\,,
\qquad\beta\!\left(\boldsymbol{k},\boldsymbol{p}\right)=\frac{
\left(\boldsymbol{k}+\boldsymbol{p}\right)^{2}\boldsymbol{k}\cdot\boldsymbol{p}}
{2\, k^{2}p^{2}}\,~.
\end{equation}
The $\Omega_{ab}$ matrix is given by,
\begin{equation}
\Omega_{ab}=\left(\begin{array}{cc}
\;\;\,0 & -1\\
-\frac{3}{2}\,\Omega_{m} & \;\;\,1+\frac{1}{\mathcal{H}^{2}}\frac{\partial\mathcal{H}}{\partial\tau}
\end{array}\right)=\left(\begin{array}{cc}
\;\;\,0 & -1\\
-\frac{3}{2} & \;\;\,\frac{1}{2}
\end{array}\right)\,.
\label{eq: Omega_ab-1}
\end{equation}
The last equality applies in an EdS background. It is straightforward to generalize $\Omega_{ab}$ to other cosmologies, such as $\Lambda$CDM \cite{Pietroni:2008jx}, or to include curvature effects \cite{Ben-Dayan:2014hsa}. In our numerical analysis, we use the EdS values for the rescaled variables $\psi=(\delta,-\theta/{\cal H} f)$, see e.g. \cite{Bernardeau:2001qr}. The error induced by this approximation is subdominant.
\vskip 4pt 

The main observation now is the following. In the limit $q \to 0$, the integral in \eqref{eq:P1N} is dominated by momenta $k \gg q$. In particular, for a physical power spectrum scaling as $P(k) \simeq k$ in the soft regime, we have, when $ k \lesssim q$,
\be
\int^{q}  \, d^3k \, P(k) \sim q^4 \, .
\ee
This is subdominant, in the $q$-soft limit, with respect to the contributions from modes with $k \gg q$, scaling as $q^2\sigma^2$, like we find in \eqref{a3}. Moreover, as we see below, $\sigma^2$ is dominated by modes near the non-linear scale.

In what follows we split the solution to the equation \eqref{eq:P1N} in three sets,
\beq 
  P_{ab}(q,\eta) = P_{ab}^{\rm hom}(q,\eta) + P^\alpha_{ab}(q,\eta) + P^\beta_{ab}(q,\eta). 
\eeq
The first contribution corresponds to the homogeneous solution, which is identical to the linear power spectrum, $ u_a u_b P_L(q,\eta)$ with $u_a=(1,1)$, when ignoring the decaying mode. The other two contributions are sourced by the $\alpha$- and $\beta$-terms, respectively. We discuss $P^\beta_{ab}$ first.

\subsubsection{$\beta$-terms}

The $\beta$ contribution is of the form (the subscript $a \in \{1,2\}$ is arbitrary)
\be
\int \frac{d^3 k}{(2\pi)^3} \, \beta(-\bk+\bq/2, \bk +\bq/2) \, B_{a22}(-\bq,\bq/2 -\bk, \bq/2+\bk) \, .
\ee
This is quite fortunate, since the vertex function scales like $\beta \propto q^2/k^2$ for $q\ll k$, and thus the integral
amounts to an angular average of the bispectrum,
\be
- \frac{q^2}{2} \int \frac{d^3 k}{(2\pi)^3}\frac{ B_{a22}(-\bq, \bq/2-\bk, \bk +\bq/2) }{k^2} \, .
\ee
This means, using the OPE, only $f_{22}^{(0)}(k,\eta)$ contributes to the above integral at leading order in~$q$. At the end of the day we obtain 
\beq
\partial_\eta \, P^{\beta}_{ab}\!\left(q,\eta\right) 
+ \Omega_{ac} P^{\beta}_{cb}\!\left(q,\eta\right) 
+ \Omega_{bc} P^{\beta}_{ac}\!\left(q,\eta\right) 
=
\frac{q^2}{2} \, P_L(q,\eta) \, 
\left(\begin{array}{cc}
0 & 1 \\
1 & 2 \\
\end{array}\right) C_{22}(\eta),
\label{a3beta}
\eeq
with
\beq
C_{22}(\eta) = 4\pi \int  dk~f^{(0)}_{22}(k,\eta),
\eeq
for the contribution to the dynamics from $\beta$-terms.\vskip 4pt
 
We could in principle directly use \eqref{EqOPE2} to extract $f_{22}^{(0)}(k,\eta)$ from the bispectrum in the soft limit. (After all, this is similar to extracting the $f^{\rm loc}_{\rm NL}$ (local non-Gaussianity) parameter in the squeezed limit for the CMB counter-part.) However, because this case involves the angular-averaged bispectrum, we can use the soft limit relations discussed in 
sec.~\ref{sec:direct}. Using \eqref{EqOPE2} and \eqref{eq:soft32}, and performing the angular integration, we recover \eqref{c22} (see also \eqref{eq:soft} and \cite{Ben-Dayan:2014hsa}.)
Moreover, using the approximation $\partial_\kappa \psi^K_2 \simeq \tfrac{4}{7}(\partial_\eta + 1)\psi^{K=0}_2$ \cite{Ben-Dayan:2014hsa}, we obtain
\be
C_{22}(\eta) \simeq \frac{12}{7} \sigma^2_{22}(\eta)+ \frac{13}{7} \frac{\partial}{\partial \eta}  \sigma^2_{22}(\eta)  \, .
\ee
We will use this result to compare against numerical simulations.

\subsubsection{ $\alpha$-terms}

While the $\alpha$-terms appear to contribute at order $q$, 
\beq
\alpha(\bk+\bq/2, \bq/2- \bk) = -\frac{\bq \cdot \bk}{k^2} + \frac{q^2}{2k^2} - \frac{(\bq \cdot \bk)^2}{q^4},
\eeq
since the integrand does not change if both $\bk$ and $\bq$ change sign, the leading
contribution will ultimately scale as $q^2$, similarly to the $\beta$-terms. Then, using \eqref{EqOPE2} we get 
\beq
\partial_\eta \, P^\alpha_{ab}\!\left(q,\eta\right) 
+ \Omega_{ac} P^\alpha_{cb}\!\left(q,\eta\right)  
+ \Omega_{bc} P^\alpha_{ac}\!\left(q,\eta\right)  
= - 
\frac{q^2}{2} \, P_L(q,\eta) \left(\begin{array}{cc}
2 & 1 \\
1 & 0 \\
\end{array}\right)C_{12}(\eta)  \, ,
\label{a3alpha}
\eeq
with
\beq
C_{12}(\eta) = C^{(0)}_{12}(\eta) + C^{(1)}_{12}(\eta) + C^{(2)}_{12}(\eta)~,
\eeq
and
\bea
C^{(0)}_{12}(\eta) &=& -\frac{4\pi}{3}\int dk~f^{(0)}_{12}(k,\eta)~, \\ 
C^{(1)}_{12}(\eta) &=& = \frac{8\pi}{3} \int dk~f^{(1)}_{12}(k,\eta)~,\\ 
C^{(2)}_{12}(\eta) &=& \frac{16\pi}{15} \int dk~f^{(2)}_{12}(k,\eta)~.
\eea
\vskip 4pt
As we discussed in sec. \ref{sec:direct}, we can then resort to the locally curved anisotropic universe to write $C_{12}(\eta)$ in terms of the variance of the displacements, $\sigma_{12}^A$, and derivatives with respect to geometrical quantities. This allowed us to infer that $C_{12}(\eta)$ does not depend significantly on modes beyond the non-linear scale. However, it requires using the Newtonian mapping for a directional mode to ${\cal O}(q^3)$. This is rather cumbersome, as we see explicitly in sec.~\ref{sec:direct2}.   Moreover, unless simulations can be performed in anisotropic backgrounds, it would be unclear how to replace derivatives with respect to these cosmological parameters by information in our universe which could be used as input in \eqref{a3}. This would be desirable in order to most accurately confront our equation with observations, or numerical methods, and asses the importance of the EFT coefficient(s) non-perturbatively. There is however another, perhaps more suggestive, way to parameterize the $\alpha$-terms. By a judicious shift in the integration variables we can re-write their contribution as follows: 

\begin{align}
\label{eq:alpha_terms2}
 \hskip -2 cm
&\int \frac{d^3k}{(2\pi)^3} \, \alpha(\bk+\bq/2, \bq/2- \bk) \, B_{a12}(\bq, \bq/2 - \bk, \bk + \bq/2) = \\
  &\int \frac{d^3k}{(2\pi)^3} \, \frac{\bq \cdot \bk}{2k^2} \,
\left[B_{a12}(-\bq, -\bq/2 - \bk, \bk - \bq/2)-B_{a12}(\bq, \bq/2 - \bk, \bk + \bq/2) \right]  \, \nn .
\end{align}
This way we see there is no angular-averaged contribution. Moreover, using an OPE for the term between brackets
it follows that only $\tilde f_{12}^{(1)}(k,\eta)$ enters in $C_{12}(\eta)$. (We used a tilde to distinguish from the direct OPE decomposition of the bispectrum.) This is nothing but a rewriting of the previous functions. However, because of the explicit angular dependence, it may suggest that the $\alpha$-terms are less important compared to
$\beta$-terms in the soft limit. Surprisingly, this turns out to be the case in standard perturbation theory. The $\alpha$-terms are roughly a factor of $20$ smaller than the leading contribution from $\beta$-terms. In the non-perturbative regime, there is no justification for neglecting these terms, which may still lead to sizable errors. However, since most of the modes that contribute to $C_{12}(\eta)$ come from near the non-linear scale, we may  expect only small deviations from the $\alpha$-terms. Hence, in what follows we work under this hypothesis, i.e. $P^\alpha_{ab}(k,\eta) \ll P^\beta_{ab}(k,\eta)$. As we shall see, this approximation fares well against numerical simulations. We will return to this point in sec.~\ref{sec:disc} and present some numerical estimates, for the error induced by neglecting the $\alpha$-terms, in appendix \ref{app:C12}.

\section{Numerical results}\label{sec:num}

We will now concentrate on the non-perturbative equation for the power spectrum in the soft limit that follows from the $\beta$-terms, together with \eqref{disc1}. In other words, we approximate $P_{ab}(q,\eta) \simeq P_L(q,\eta) + P^\beta_{ab}(q,\eta)$, such that
\bea
\partial_\eta \, P_{ab}\!\left(q,\eta\right) &=& 
-\Omega_{ac} P_{cb}\!\left(q,\eta\right)
- \Omega_{bc} P_{ac}\!\left(q,\eta\right) \nn \\
&& -  
q^2 \, P_L(q,\eta) \, 
\left(\begin{array}{cc}
0 & 1 \\
1 & 2 \\
\end{array}\right)
\left( \frac{6}{7} \sigma^2_{22}(\eta)
+ \frac{13}{14} \partial_\eta  \sigma^2_{22}(\eta) \right) \, .
\label{eq:flow_final}
\eea

\begin{figure}[h!]
\centering
\includegraphics[width=0.5\textwidth]{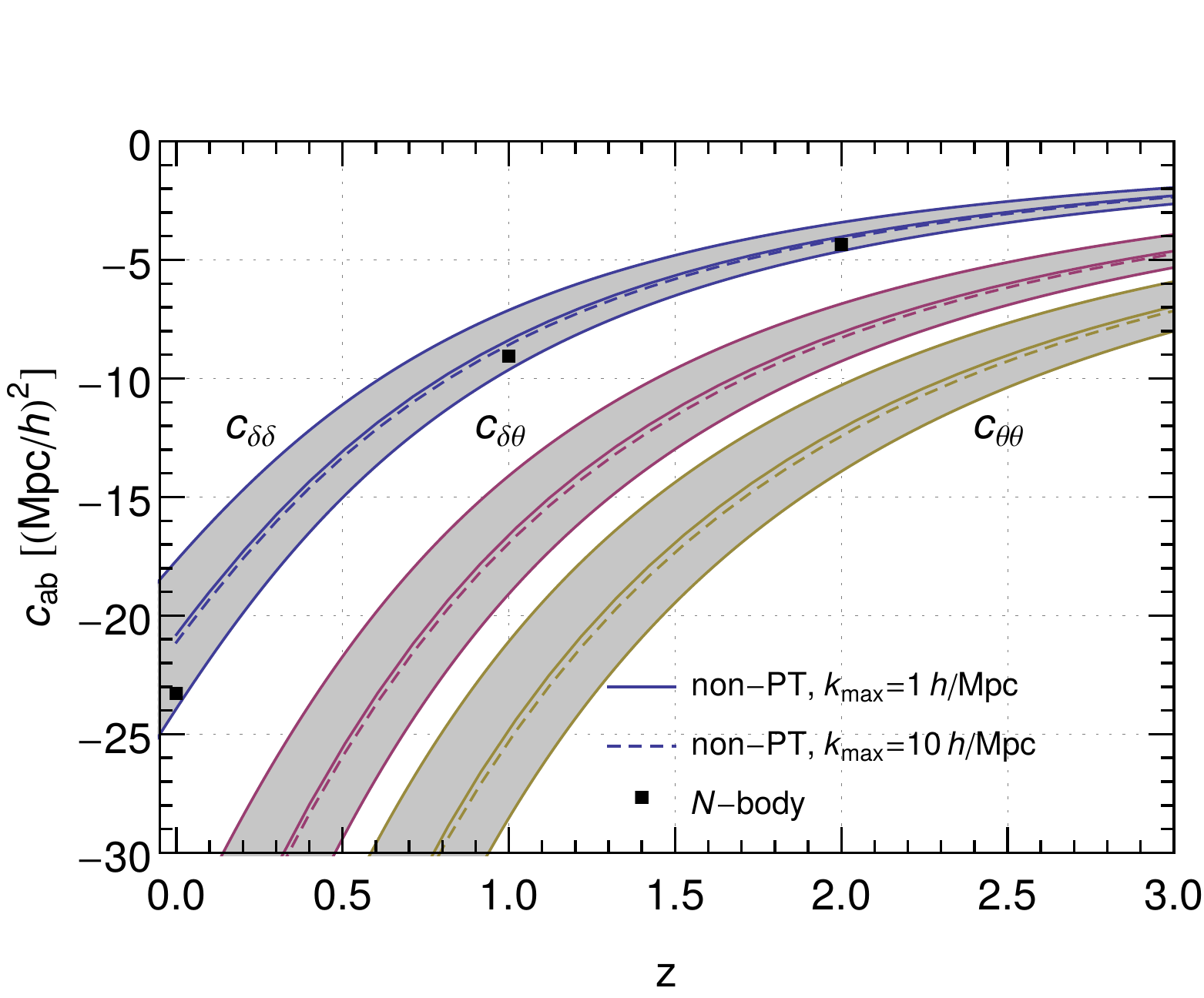}
\caption{Numerical results for for the coefficients $c_{11}=c_{\delta\delta}, c_{12}=c_{\delta\theta}, c_{22}=c_{\theta\theta}$ as a function of redshift, obtained from the (truncated) non-perturbative equation \eqref{eq:flow_final}, using velocity correlation spectra from $N$-body simulations performed in \cite{Hahn:2014lca} as input. The black squares in the plot correspond to numerical results for the $c_{\delta\delta}$ coefficient for the density power spectrum in the soft limit \cite{Foreman:2015, Baldauf:2015aha}. The error bands are a rough estimate --at the $\sim 10 \%$ level-- of the errors in the non-perturbative equation, induced by ignoring $\alpha$-terms and using the approximation in \eqref{disc1}. The agreement is remarkably good, especially at higher redshifts.}
\label{fig:sig3n}
\end{figure}

Furthermore, we parameterize the power spectrum in the soft limit by 
\beq 
P_{ab}(q,\eta)\to P_L(q,\eta)\Big(1+c_{ab}(\eta)q^2\Big),
\label{eq:cabDef}
\eeq 
and, after subtracting the linear contribution, the above equation becomes a differential equation for the coefficients $c_{ab}(\eta)$.\vskip 4pt 

A simple exercise is to use the linear variance $\sigma_{22,lin}^{2}\equiv \sigma_{lin}^{2}\propto e^{2\eta}$
on the right-hand side, which yields at leading-order 
\be\label{eq:cflowLO}
 c^{\rm  LO}_{ab}(\eta) = \frac{\sigma_{lin}^2(\eta)}{63} \left(\begin{array}{cc} -38 & -76 \\ -76 & -114\end{array}\right) 
\simeq \sigma_{lin}^2 (\eta)\left(\begin{array}{cc} -0.60 & -1.21\\ -1.21 & -1.81\end{array}\right)\;.
\ee
This must be compared to the direct standard perturbation theory one-loop computation, which gives
\be\label{eq:cspt1L}
 c^{\rm SPT}_{ab}(\eta) = \frac{\sigma_{lin}^2(\eta)}{105} \left(\begin{array}{cc} -61 & -125 \\ -125 & -189\end{array}\right) 
\simeq \sigma_{lin}^2 (\eta) \left(\begin{array}{cc} -0.58 & -1.19\\ -1.19 & -1.80\end{array}\right)\;.
\ee

\begin{figure}[h!]
\begin{center}
\includegraphics[width=0.4\textwidth]{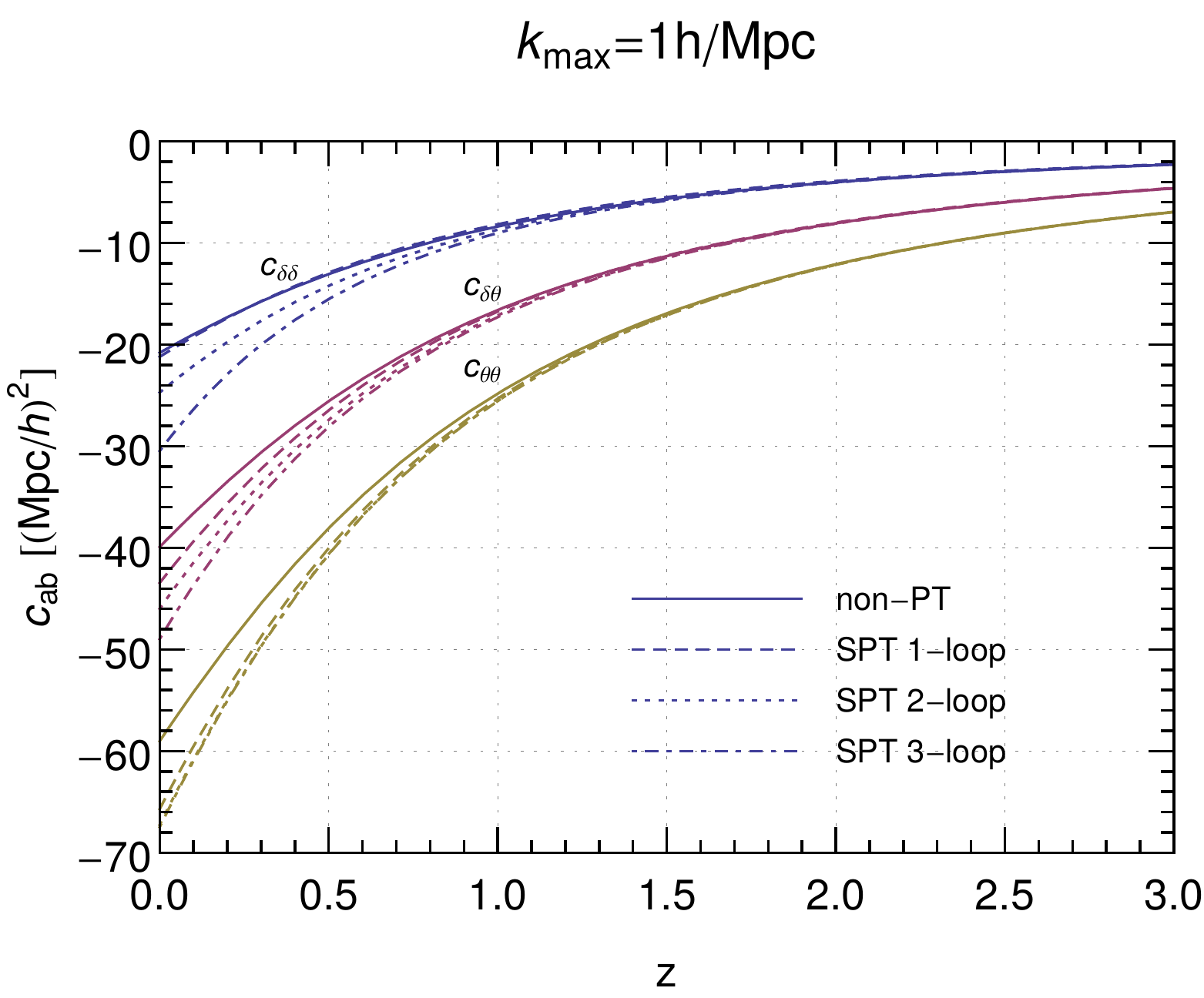}
\includegraphics[width=0.4\textwidth]{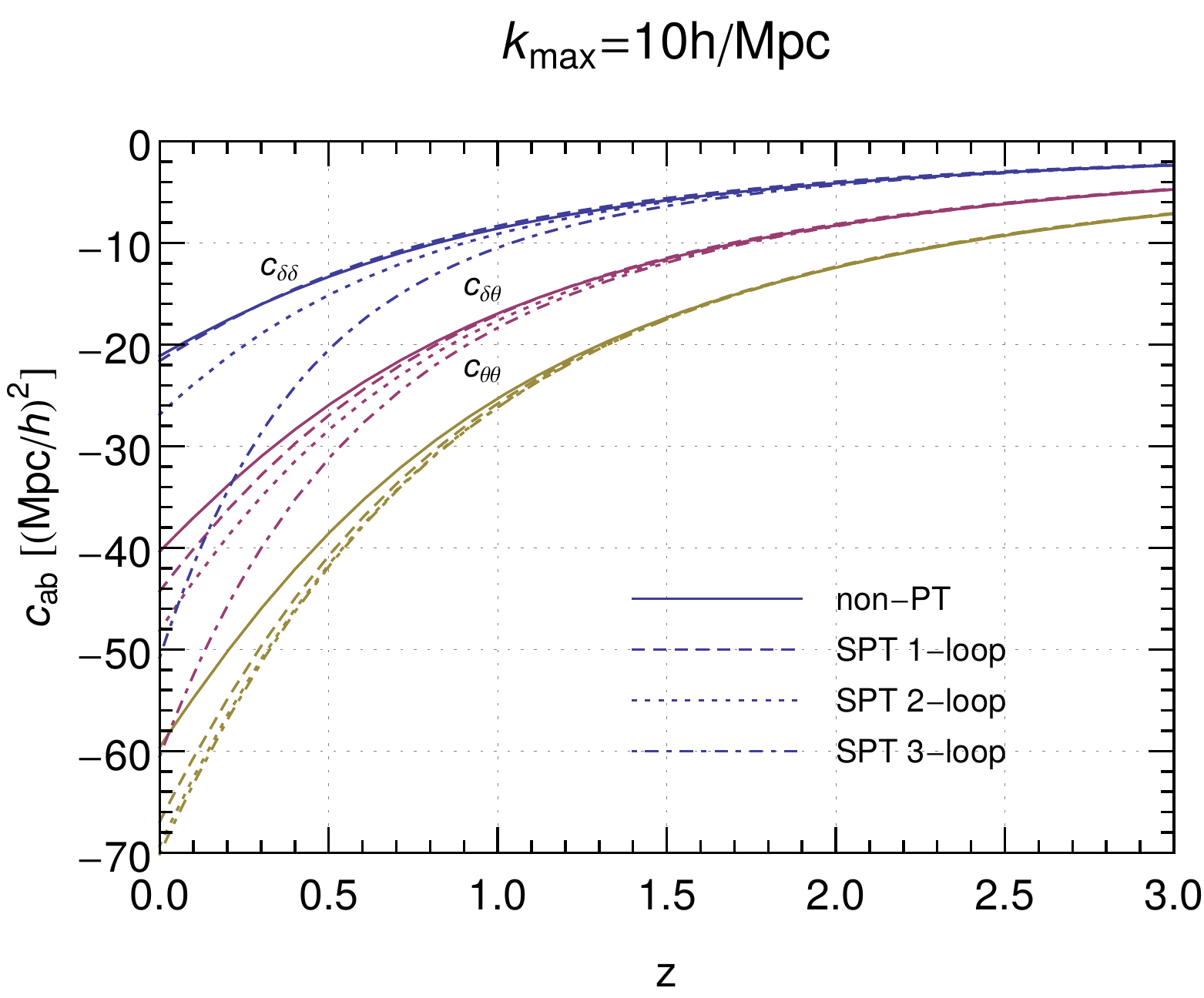}
\includegraphics[width=0.4\textwidth]{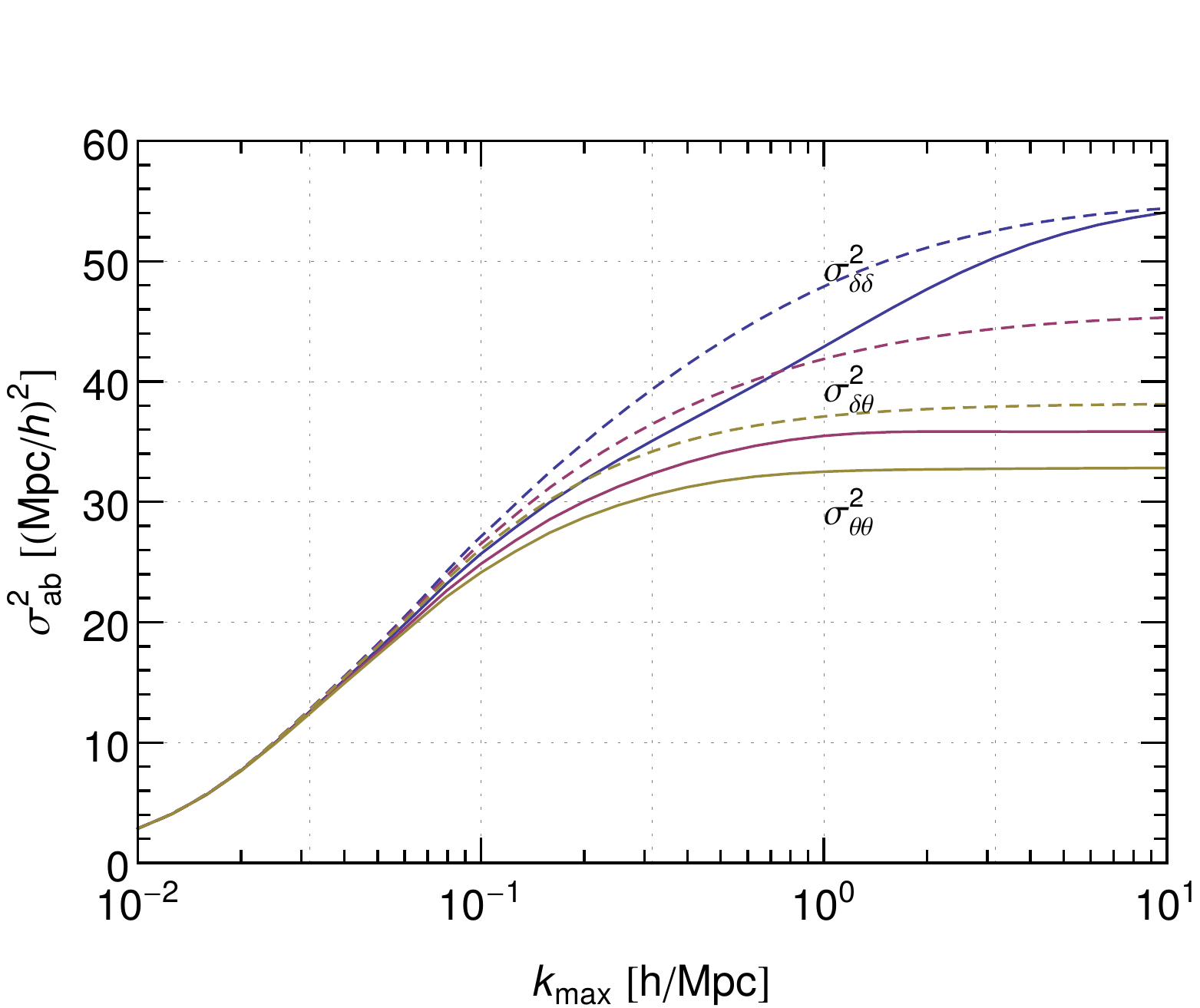}
\includegraphics[width=0.4\textwidth]{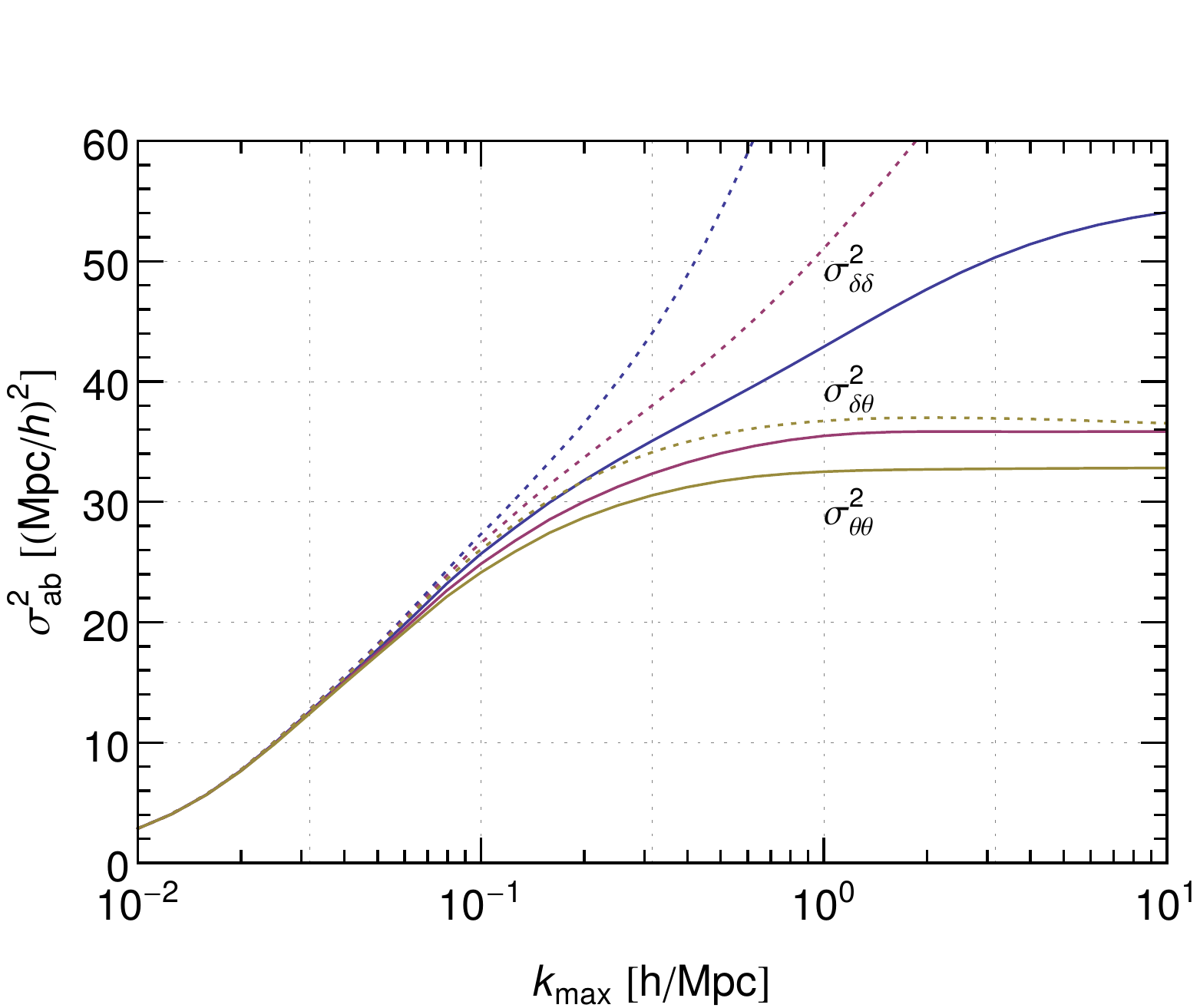}
\end{center}
\centering\caption{On the upper panels, solid lines reproduce the results from Fig.~\ref{fig:sig3n}. Dashed lines show the one-loop, dotted lines the two-loop, and dot-dashed lines the three-loop standard perturbation theory results. The two figures are given for two different cutoff scales $k_{max}=1(10)\,h/$Mpc, respectively. The solid lines are affected at the $\simeq 2\%$ level by  the change in cutoff. For the lower panels, on the left we plot $\sigma_{ab}^2(k_{\rm max})$ (in dashed lines) computed at one-loop as a function of cutoff. On the right, we plot (in dotted lines) the two-loop calculation. We notice, once again, the relatively strong dependence on $k_{\rm max}$ beyond the non-linear scale. This is the case also for $\sigma^2_{12}(k_{\rm max})$, especially at two-loops, and also for $\sigma^2_{22}(k_{\rm max})$ to a lesser degree. On the other hand, the non-perturbative results remain essentially unaltered.}
\label{fig:ksqCoeff}
\end{figure}

At this level, the small difference quantifies the influence of the $\alpha$-terms in the flow equation, we alluded before, as well as the error in the approximation in \eqref{disc1}, which we argued in \cite{Ben-Dayan:2014hsa} is small within perturbation theory. For the non-linear regime we estimated the error to be at the $\sim 10\%$ level, which is consistent with Fig.~\ref{fig:sig3n}. (See also the next section.)
The input from UV modes in the evolution equation enters through $\sigma_{22}^2(k_{\rm max})$, the variance of the velocity divergence auto correlation up to a given cutoff. To extract this variance we use results from $N$-body simulations based on a phase-space projection technique \cite{Hahn:2014lca}. The result is given in Fig.~\ref{fig:sigma}, also shown in the lower panels in Fig.~\ref{fig:ksqCoeff}. For the numerical results we used $\Lambda$CDM model with $\Omega_m=0.276$, $\Omega_\Lambda=0.724$, $\Omega_b=0.045$, $h=0.703$, $\sigma_8=0.811$, $n_s=0.96$. The value of the $c_{ab}(\eta)$ coefficients as a function of redshift is plotted in Fig.~\ref{fig:sig3n}. For the density coefficients, the agreement is remarkably good within the error bars. Unfortunately, we do not currently have knowledge of data we could use to compare our predictions for the $c_{22}$ and $c_{12}$ coefficients.
\vskip 4pt

In order to explore the cutoff dependence in the UV parameters, in the top panels in Fig.~\ref{fig:ksqCoeff} we show the numerical and perturbative results obtained using two different cutoff scales $k_{max} = 1(10)\,h/$Mpc. While the perturbative computation depends strongly on the cutoff at the two- and three-loop level, the full solution is essentially robust. For the non-perturbative solution, the $c_{ab}$ coefficients do not vary significantly when the cutoff changes by an order of magnitude, and the main contribution comes from modes with $k \lesssim 1\,h/$Mpc. 
The apparent strong(er) dependence on small-scale fluctuations indicated in perturbative calculations, displayed in the standard two- and three-loop results, is an artifact. This is also exemplified in the lower panels of Fig.~\ref{fig:ksqCoeff}, where we see the dependence of the variance $\sigma_{ab}^2(k_{\rm max})$ on $k_{\rm max}$, within standard one- and two-loop perturbation theory, contrasted with the non-perturbative results.

%
\section{Discussion} 
\label{sec:disc}
%

We derived a non-perturbative relation for the power spectrum in the limit of long wavelength perturbations $q\to 0$,
\bea
\partial_\eta \, P_{ab}\!\left(q,\eta\right) &=& -\Omega_{ac} P_{cb}\!\left(q,\eta\right) -
\Omega_{bc} P_{ac}\!\left(q,\eta\right) \nn \\ 
&& \hskip -3 cm
- \frac{q^2}{2} \, P_L(q,\eta) \, 
\Bigg\{ \left(\begin{array}{cc}
0 & 1 \\
1 & 2 \\
\end{array}\right) C_{22}(\eta)
+
\left(\begin{array}{cc}
2 & 1 \\
1 & 0 \\
\end{array}\right)C_{12}(\eta) \Bigg\} \, ,
\label{a3disc}
\eea
with coefficients $C_{22(12)}(\eta)$ which depend on the variance of velocity-velocity and velocity-density correlations on short scales, albeit in a spatially curved anisotropic universe. Let us emphasize, once again, that the density--density correlation on short scales does not enter in \eqref{a3disc}, due to the structure of the non-linear couplings $\gamma_{abc}$. The main ingredients are the OPE together with relations in the soft limit between the three-point function and the power spectrum in a modified cosmology. \vskip 4pt

The presented derivation has important ramifications. Perhaps the most relevant concerns the UV dependence of the power spectrum in the soft limit, and the implications for the EFT of LSS.  At leading order in the soft wavenumber $q$ the EFT of LSS introduces a new parameter, $l_{\rm ren}^2$, which scales as $q^2 P_L(q,\eta)$ when $q \to 0$. This coefficient includes a series of response functions in LEFT, discussed in \cite{left}, or may be interpreted as a `sound speed', $c_s^2$, in the Eulerian approach~\cite{eft2}. As it was originally emphasized in \cite{Baumann:2010tm,eft2,left}, the effects of the EFT coefficient(s) can be read off from the discrepancy between observations, $P_{\rm obs}(k,\eta)$, and the $P_{\rm flow}(k,\eta)$ which results from solving \eqref{flow} without the addition of finite size effects. Alternatively, we can use our non-perturbative equation in \eqref{a3disc} 
to find the solution for the power spectrum. In order to assess the dependence of $l_{\rm ren}^2$ on the short-distance modes, we introduced a cutoff scale, $k_{\rm max}$. This is the maximum wavenumber contributing to the variance of the displacements in density and velocity fields. The dependence on $k_{\rm max}$ in \eqref{a3disc} enters through the coefficients $C_{22(12)}(\eta)$, which are determined as a function of the variance $\sigma_{22(12)}^A(k_{\rm max},\cdots)$ in a (hypothetical) locally curved anisotropic background cosmology. Hence, for a given $k_{\rm max}$, we can define a (renormalized) parameter which tracks the dependence on short scales as follows: 
\beq
\label{lrenkmax}
 l_{\rm ren}^2(k_{\rm max}) \equiv \frac{\Delta P(q,k_{\rm max})}{q^2 P_L(q)},~~~ q \ll \Lambda_{\rm NL},
\eeq
at a given time. The observed $l^2_{\rm ren}$ is given by taking the cutoff to infinity, or in practice $k_{\rm max} \gg \Lambda_{\rm NL}$, i.e. much bigger than the non-linear scale.  The dependence on $k_{\rm max}$ then reflects the sensitivity to short-distance modes, whereas the importance of these coefficients depends on $\Delta P(k) \equiv P_{\rm obs}(k) - P_{\rm flow}(k,k_{\rm max})$. Namely, the mismatch between observations (and/or simulations) and solutions to \eqref{flow}.\vskip 4pt In principle the mismatch, $\Delta P(k)$, also measures the effect of vorticity, ${\bf \boldsymbol{w}} =\nabla \times {\bv}$, which was ignored in \eqref{a3disc}. The vorticity enters in the density power spectrum through new vertices in \eqref{flow}, e.g. $\gamma_{\delta\omega\theta}$, and in turn these contribute to the coefficients in \eqref{a3disc} through terms proportional to e.g. $\sigma^2_{\omega\omega}(k_{\rm max})$. However,  $P_{\omega\omega}(k)$ is highly suppressed on long-distance scales, and moreover, on short scales it is at most of the same order as the velocity power spectrum~\cite{Hahn:2014lca}. Hence, we conclude $\sigma^2_{\omega\omega}(k_{\rm max}) \ll \sigma^2_{22}(k_{\rm max})$.\vskip 4pt 

\begin{figure}[t!]
\centering
\includegraphics[width=0.4\textwidth]{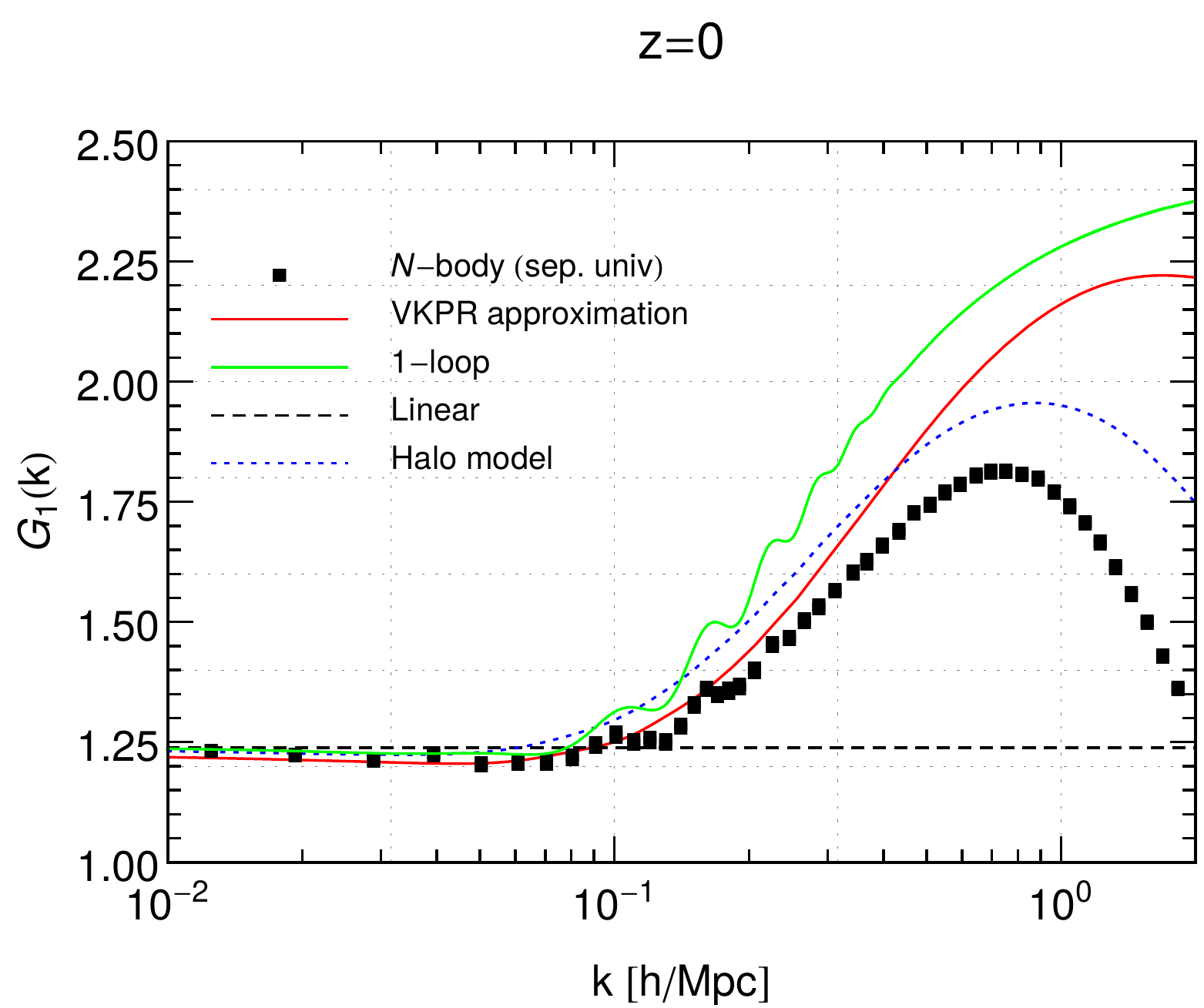}
\includegraphics[width=0.4\textwidth]{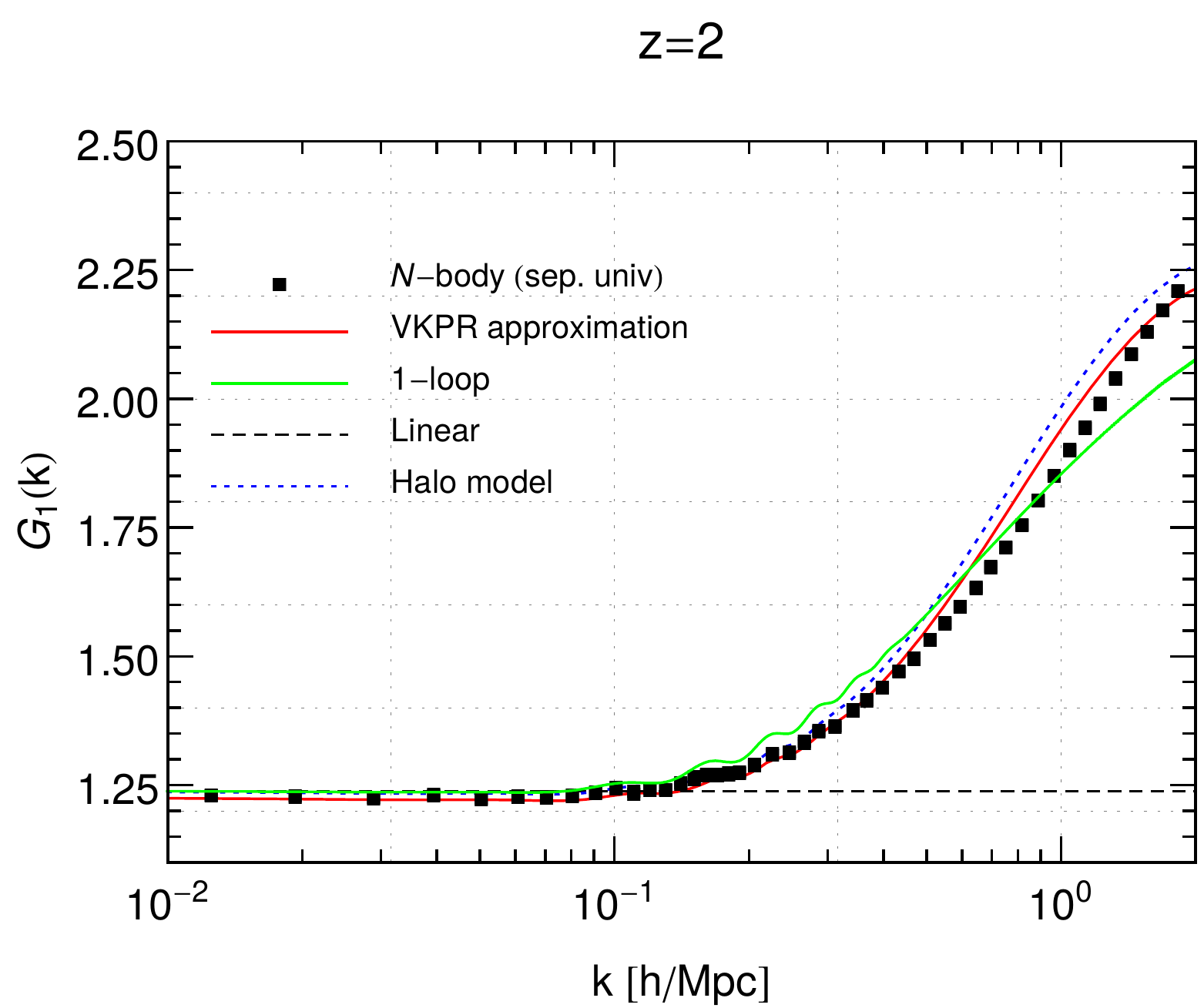}
\caption{The response of the power spectrum to a long-wavelength perturbation, $G_1(k,z)$, at redshifts $z=0$ (left) and $z=2$ (right), as introduced in \cite{Wagner:2015gva}. The red solid line uses the approximation in \eqref{vkpr} \cite{Valageas:2013zda,Kehagias:2013paa} (see footnote \ref{fn:G1}). The results from the $N$-body simulations are reproduced here from~\cite{Wagner:2015gva} by the black squares. The other curves represent perturbative computations and the halo model. We see relatively good agreement between the approximation in \eqref{vkpr} and $N$-body results at low $k$'s, as it was also pointed out in \cite{Ben-Dayan:2014hsa}. However, \eqref{vkpr} fails for high $k$'s at zero redshift, with an error of order $\sim 10\%$ near the non-linear scale. The disagreement becomes more dramatic as we increase $k$, with the numerical result much below the red line for $k \gg \Lambda_{\rm NL}$.}
 \label{G1z0}
\end{figure}

\begin{figure}[t!]
\centering
\includegraphics[width=0.55\textwidth]{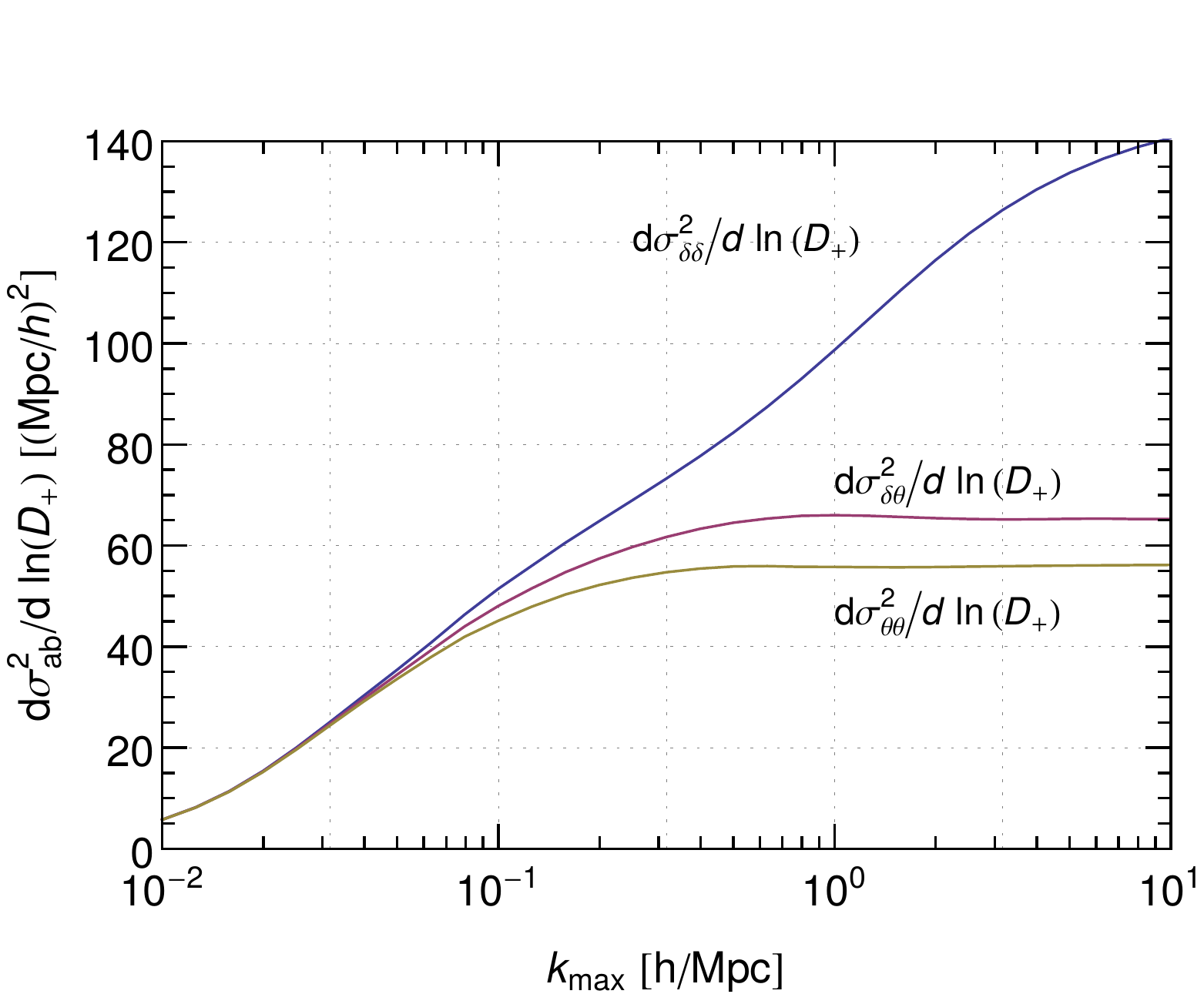}
\caption{Derivatives of the moments introduced in \eqref{eq:sig3} as a function of $k_{max}$ and at zero redshift. Results based on $N$-body simulations presented in \cite{Hahn:2014lca}. Notice that $\partial_\eta \log \sigma^2_{22(12)}(k_{\rm max}) \simeq 2$ for $k_{\rm max} \gtrsim \Lambda_{\rm NL}$. (See Fig.~\ref{fig:sigma}.)}
 \label{dsigma}
\end{figure}

The $\sigma_{ab}^2(k_{\rm max})$ are plotted in Fig.~\ref{fig:sigma} as a function of the cutoff, in our universe. Notably, $\sigma^2_{22}(k_{\rm max})$ and
$\sigma^2_{12}(k_{\rm max})$ show a saturation on short scales. On the other hand, $\sigma^2_{11}(k_{\rm max})$ exhibits a
different behavior. As we argued, the qualitative behavior on very short scales should not depend strongly on details of the background cosmology, which are only relevant at long distances. For example --if we concentrate on the curvature dependence for a spherically symmetric perturbation-- we expect 
\beq
\sigma_{ab}^K(k_{\rm max},\eta) \simeq \sigma_{ab}(k_{\rm max},\eta),
\eeq
beyond the non-linear scale. Moreover, we also expect the power spectrum on short scales to obey ($\kappa = K/(a^2H^2)$) 
\beq  
\left. \frac{\partial}{\partial \kappa} P_{ab}^K(k,\eta)\right|_{K=0} < P_{ab}(k,\eta), \label{former}
\eeq
for $k \gg \Lambda_{\rm NL}$. This can be shown to be the case from numerical simulations \cite{Li:2014sga,Wagner:2014aka,Chiang:2014oga,Wagner:2015gva}. For instance, in \cite{Wagner:2015gva} the derivative with respect to curvature was calculated, and used to infer the response to a long-wavelength perturbation using the separate universe approach. We reproduce their results in Fig.~\ref{G1z0} for the case of the density power spectrum. As we see in the plot, the numerical result lies {\it below} the red solid curve, which is constructed using the approximation in \eqref{disc1}, e.g. $\partial_\kappa \psi_1^K \simeq \frac{4}{7} \partial_\eta \psi^K_1$.\footnote{\label{fn:G1} Strictly speaking the authors in \cite{Wagner:2015gva} evaluated numerically the response function: 
\beq G_1(k,\eta) \equiv -\frac{1}{3}\partial_\eta \log P_{11}^{K=0}(k,\eta) +  \frac{5}{3} \left.\partial_\kappa \log P_{11}^K(k,\eta)\right|_{K=0},\eeq for the density power spectrum at different redshifts. This is the combination which enters in \eqref{eq:soft}. In Fig. \ref{G1z0}, we compare the results reported in \cite{Wagner:2015gva} with the (VKPR) approximation implicitly used in \cite{Valageas:2013zda,Kehagias:2013paa}, see also \cite{Ben-Dayan:2014hsa}, \beq \label{vkpr} \left.\partial_\kappa \log P_{11}^K(k,\eta)\right|_{K=0} \to \frac{4}{7}\partial_\eta \log P_{11}^{K=0}(k,\eta)~.\eeq }
The error is of order $\sim 10\%$ around $k \simeq 0.5 \,h/$Mpc at redshift zero, and the separation grows as we increase~$k$. While the results in \cite{Wagner:2015gva} did not include velocity correlators, we expect similar results.
On the other hand, we plot $\partial_\eta \sigma^2_{ab}(k_{\rm max})$ as a function of $k_{\rm max}$ in Fig.~\ref{dsigma}.
We notice, first of all, the weak dependence on hard modes beyond the non-linear scale for the $22$- and $12$-component and, moreover, the approximation $\partial_\eta \log \sigma^2_{22(12)}(k_{\rm max}) \simeq 2$ is relatively accurate for $k_{\rm max} \gtrsim \Lambda_{\rm NL}$. From these observations we conclude that \eqref{former} is fulfilled. Therefore, the weak sensitivity to hard modes beyond the non-linear scale also applies for the derivative of $\sigma^K_{22(12)}$ with respect to spatial curvature, which is what enters in \eqref{a3disc}, e.g. \eqref{c22}. On physical grounds we expect a similar behavior in an anisotropic universe, and the contributions to \eqref{c12}, for both the variance and the variation with respect to the extra geometrical parameters introduced in sec.~\ref{sec:direct2}. \vskip 4pt

From the above reasoning, we concluded that the $C_{22(12)}(\eta)$ vary weakly with scales beyond the non-linear regime. This implies
\beq P_{\rm flow} (q,k_{\rm max}) \simeq P_{\rm flow} (q,\Lambda_{\rm NL})~,~~~k_{\rm max} \gg \Lambda_{\rm NL}~, \eeq
and in turn, for the parameter introduced in \eqref{lrenkmax},
\beq
 l_{\rm ren}^2 (k_{\rm max} \gg \Lambda_{\rm NL}) 
 \simeq \frac{\Delta P(q,\Lambda_{\rm NL})}{q^2 P_L(q)}~,~~~ q \ll \Lambda_{\rm NL}~.
\eeq
 
Hence, the leading order renormalized coefficient(s) in the EFT framework themselves may be obtained with information from modes up-to/near the non-linear scale.\vskip 4pt 

This result is not surprising, and resonates with the statement that virialized scales decouple on large scale dynamics. This was shown in \cite{Baumann:2010tm} for the back-reaction on the evolution of the universe, since short scales contribute through the expectation value of their quadrupole moment~\cite{left}, which is suppressed relative to non-virialized scales \cite{Peebles:1980}. Likewise, our analysis demonstrates the same occurs for the response of this quadrupole to a long-wavelength perturbation, which enters at order $q^2$.\footnote{Notice that for virialized objects $\delta_{\rm vir} \simeq 10^2$. However, using that the potential is roughly constant (of order $\Phi \simeq 10^{-5}$) on all scales, this corresponds to $k_{\rm vir} \simeq 10\,\Lambda_{\rm NL}$. Furthermore, the power spectrum turns over near $ k\simeq k_{\rm eq} < \Lambda_{\rm NL}$, namely the horizon scale at matter/radiation equality. This introduces an extra suppressing factor, and is the reason we see the turn in Fig.~\ref{fig:sigma} closer to $\Lambda_{\rm NL} \simeq 0.5~h/$Mpc.} Our results are then compatible with various numerical investigations of the impact of UV modes \cite{Pueblas:2008uv,Nishimichi:2014rra}, as well as general expectations from analytical arguments, e.g. \cite{Peebles:1980,Rees:1982,Goroff:1986ep}.\vskip 4pt

This is, on the other hand, in stark contrast with what occurs when standard Eulerian (or Lagrangian) perturbation theory 
is used to solve for \eqref{flow}, where $P_{\rm flow}(k,\eta)$ is calculated as an expansion in the linear power spectrum, involving integrals over all scales. Within a perturbative scheme, the EFT parameters are split into a counter-term and a renormalized piece. We find that any dependence of the leading order EFT coefficient(s) on short-distance modes beyond the non-linear scales develops mainly from the counter-term(s). In other words, the UV sensitivity is nothing but an artifact due to the inapplicability of perturbative techniques beyond the non-linear scale, as opposite to intrinsic UV dependence of the --physical-- renormalized parameters encoding finite size effects. We see this in detail in sec.~\ref{sec:num}, where we observe in Fig.~\ref{fig:ksqCoeff} the cutoff dependence in loop computations, in contrast with the non-perturbative results. \vskip 4pt 

The fact that modes beyond the non-linear scale do not have a major influence on the renormalized parameter(s) does not mean that the effect of the new term in the dynamics of \eqref{a3disc}, $l_{\rm ren}^2 q^2 P_L(q)$, is necessarily small. In fact, its relevance relies on how well the solution to \eqref{flow} fares against observations. Moreover, on physical grounds, we know that {\it tidal effects} are expected to contribute on long scales \cite{left,nrgr}. In order to estimate the size of the contribution from the renormalized EFT parameter(s) we used a truncated version of our non-perturbative equation.  Our truncation sets $C_{12}(\eta)=0$ on the right-hand-side of \eqref{a3disc}, and uses the approximation $\partial_\kappa \psi^K_2 \simeq \tfrac{4}{7}(\partial_\eta+1)\psi^{K=0}_2$ applied to \eqref{c22}, which implies 
\beq
\label{c22disc}
C_{22}(\eta)= \frac{1}{7}\left( 12~\sigma^2_{22}(\eta) + 13 \frac{\partial}{\partial \eta}\sigma^2_{22}(\eta)\right).
\eeq
It is debatable to what extent this expression is valid in the non-perturbative regime. As we showed, due to specific features of the equations in \eqref{flow}, it seems to be remarkably accurate --percent level-- in perturbative computations \cite{Ben-Dayan:2014hsa}. However, confronted with $N$-body simulations for the standard $\Lambda$CDM cosmology the accuracy degrades on shorter scales to $\sim 10\%$ at $z=0$ and $k \simeq 0.5 \,h/$Mpc, see Fig.~\ref{G1z0} \cite{Li:2014sga,Wagner:2014aka,Chiang:2014oga,Wagner:2015gva}.
This is indeed expected, as we argued the derivatives with respect to curvature and time should not be interchangeable beyond the non-linear scale, see \eqref{former}. However, since the modes which contribute to $C_{22}(\eta)$ the most live near the non-linear scale, the approximation in \eqref{c22disc} may be more reliable. We see this explicitly in Fig.~\ref{G1z0}.\vskip 4pt In addition, we also have the $\alpha$-terms encoded in $C_{12}(\eta)$, which we neglected in our simplified equation. These are subleading within perturbation theory --a factor of $20$ smaller-- and their relevance is unclear in the non-perturbative regime.  Since, as we argued, only modes near $k_{\rm max} \simeq \Lambda_{\rm NL}$ contribute to these coefficients, we do not expect the $\alpha$-terms to modify our results radically. Hence, we estimated overall an error in our truncation roughly of order $\sim 10 \%$ at $z=0$. This level of precision is consistent with the results presented in sec.~\ref{sec:num}, see Fig.~\ref{fig:sig3n}, and also with a simple estimate of the impact of $C_{12}(\eta)$ in appendix~\ref{app:C12}. For example, our non-perturbative (restricted) solution for the density power spectrum, given by (in redshift space) 
\beq \label{eq5} P_{\delta\delta}(q,z) = (1+ c_{\delta\delta}(z)q^2) P_L(q,z)~,\eeq with \beq c_{\delta\delta}(z=0) \simeq -21~({\rm Mpc}/h)^2~,\eeq is within the error of the numerical value \cite{Foreman:2015,Baldauf:2015aha} 
\beq c_{\delta\delta}^{\rm num}(z=0) \simeq - 23~({\rm Mpc}/h)^2~.\eeq The accuracy improves the higher the redshifts. Because of the series of approximations, it is not possible to directly extract the size of the renormalized EFT coefficient(s) in the soft limit. We can, nonetheless, conclude that their importance in correcting \eqref{flow} is at most of the same order as our error bands, at the $\sim 10 \%$ level. \vskip 4pt The overall level of accuracy could be significantly improved by fitting the time dependence of the response function to a long-wavelength perturbation, $G_1(k,z)$ in Fig.~\ref{G1z0}, integrated over wavenumbers. This is ultimately what contributes to $C_{22}(\eta)$. Because of the missing $C_{12}(\eta)$ term in our truncated equation, this was not justified at this stage. (In fact, the approximation in \eqref{disc1} proved to be relatively accurate to compute $C_{22}(\eta)$, with an error which is similar to the one induced from ignoring $\alpha$-terms.) However, in principle we could also incorporate $C_{12}(\eta)$, provided response functions to directional long-wavelength perturbations could be obtained from numerical simulations, using an anisotropic separate universe approach. Contrary to a direct extraction of the $c_{\delta\delta}(z)$ coefficient in \eqref{eq5} from $N$-body simulations, where one needs large volumes to beat variance in the soft limit, e.g. \cite{Foreman:2015,Baldauf:2015aha}, computing the response functions require modest simulation volumes \cite{Li:2014sga,Wagner:2014aka,Chiang:2014oga,Wagner:2015gva}. This suggest a hybrid analytic/numerical approach to model the power spectrum. The numerical input would come from fitting the time behavior of integrated response functions, unlike extracting the low-$q$ behavior of the power spectrum. In principle, this would allow us to precisely determine the size of the renormalized leading order coefficient(s) in the EFT of LSS.\footnote{There is a technical point worth mentioning. The renormalized parameters in the EFT are not necessarily constant for all values of $q$, but may also depend on scale. The origin of this inherited $q$-dependence is the existence of non-analyticities such as logarithms, e.g. $q^2(\log q)^n$. This means, in practice, that the renormalized parameters at different scales may differ by large logarithms which need to be resumed. This is achieved through the renormalization group flow. On the one hand, our conclusion will not change when we vary $q$ towards the non-linear scale: For $q \simeq \Lambda_{\rm NL}$ the renormalized parameter(s) will not depend strongly on short-distance modes beyond the non-linear scale. On the other hand, there could still be an important flow between modes with $q \ll \Lambda_{\rm NL}$ and the non-linear scale. We do not expect this scale dependence to have a large impact, but it is nonetheless an important factor when hunting for percent-level accuracy. See \cite{Baldauf:2015aha, Foreman:2015} for a somewhat related discussion.} This is left for future work.

\newpage

\section*{Acknowledgements} 

We thank Dan Green, Enrico Pajer, Fabian Schmidt and Matias Zaldarriaga for helpful comments and discussions. We also thank Chi-Ting Chiang,  Eiichiro Komatsu, Fabian Schmidt and Christian Wagner, for sharing their numerical data in Ref. \cite{Wagner:2015gva}. R.A.P.~thanks the theory group at DESY (Hamburg), the KITP (Santa Barbara) and the ICTP (Trieste), for their hospitality while this work was being completed. L.S. and T.K. are supported by the German Science Foundation (DFG) within the Collaborative Research Center (SFB) 676 `Particles, Strings and the Early Universe.' R.A.P. is supported by the Simons Foundation and the S\~ao Paulo Research Foundation (FAPESP), under grants 2014/25212-3 and 2014/10748-5.

\appendix
\section{Estimating the impact of $C_{12}$}\label{app:C12}

In the following we present an estimate for the impact of the coefficient  $C_{12}(\eta)$ in (\ref{a3disc}), through the contribution from $P^\alpha_{ab}(q,\eta)$ to the power spectrum introduced in sec.\,\ref{sec:softfluid}. 
\begin{figure}[t!]
\centering
\includegraphics[width=0.55\textwidth]{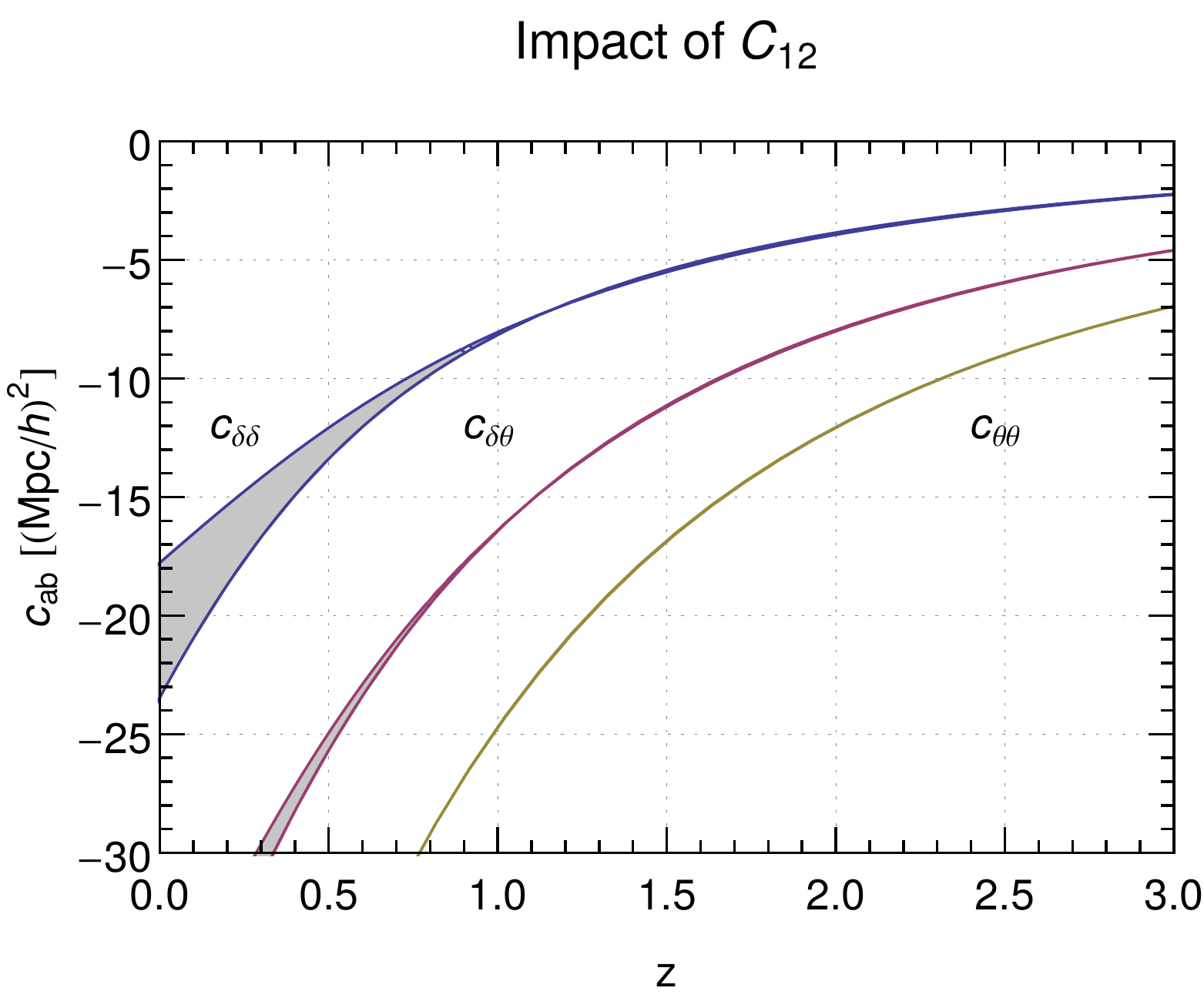}
\caption{Estimate for the impact  of $C_{12}(\eta)$ on the coefficients $c_{ab}$ in \,(\ref{eq:cabDef}). For definiteness, we used $|d_1|, |d_2| \leq 3$.}
 \label{ImpactofC12}
\end{figure}  
As was discussed in sec. \ref{sec:PS}, the $C_{12}(\eta)$ coefficient depends on the response of the correlation between velocity and density to various geometric parameters, including background curvature, etc. These response functions could in principle be extracted from suitable $N$-body simulations. In absence of such data, and inspired by \eqref{c22disc}, we then explore the following ansatz
\beq
  C_{12}(\eta)=   d_1 \, \sigma^2_{12}(\eta) + d_2 \, \frac{\partial}{\partial \eta}\sigma^2_{12} \,,
\eeq
with some adjustable free parameters $(d_1, d_2)$. In order to estimate their plausible values, we evaluate \eqref{a3disc} within perturbation theory. At leading order, $\sigma^2_{ab}\to \sigma^2_{lin}\propto e^{2\eta}$. Requiring that the left-hand side of (\ref{a3disc}) reproduces the one-loop correction to the power spectrum, and using (\ref{c22disc}), one finds
\beq
\label{ab5}
d_1+2d_2 = -\frac{2}{35}\;.
\eeq
Notice this is a significant suppression which, compared to (\ref{eq:flow_final}), precisely accounts for the $\%$-level difference between (\ref{eq:cspt1L}) and (\ref{eq:cflowLO}).\vskip 4pt 
We can then estimate the impact of $C_{12}(\eta)$ by varying $d_1$ and $d_2$ while taking \eqref{ab5} into account. The results are shown in Fig.\,\ref{ImpactofC12}.
The dependence is relatively mild, at the level of \beq \Delta c_{\delta\delta} \sim 15\%\,(5\%)\,,~~\Delta c_{\delta\theta} \sim 5\%\,(1\%)\,,~~ \Delta c_{\theta\theta} \lesssim 1\%\,,\eeq  at $z=0\, (0.5)$ respectively.\vskip 4pt
Notice that, for the $\sigma_{12}^2(\eta)$ extracted from the $N$-body simulations, one finds $\partial_\eta\ln\sigma_{12}^2\simeq 2$ at all redshifts. Hence, the influence of the $C_{12}(\eta)$ term, and related uncertainties, are then effectively suppressed due to the small overall coefficient in \eqref{ab5}. We also find that velocity components are even less affected by $C_{12}(\eta)$, which can be understood from the structure of (\ref{a3disc}). Note that the results presented in Fig.\,\ref{fig:sig3n}, which are obtained neglecting $C_{12}(\eta)$, are also within the uncertainty bands shown in Fig.\,\ref{ImpactofC12}. In addition, one can adjust the remaining free parameter in \eqref{ab5}, such that the expression in \,(\ref{a3disc}) is fulfilled (approximately) at next-to leading order in perturbation theory. In particular, the next-to-leading corrections are reproduced within a relative accuracy of $\sim 15\%$ or better for all density- and velocity- correlations (and crossed components), independently of the linear input. Since the next-to-leading contribution is suppressed compared to the leading order, by a factor of a few in the relevant regime, this level of accuracy is also compatible with our estimated uncertainty in our truncated equation (of order $\sim 10\%$, as discussed in the text.)

\addcontentsline{toc}{section}{References}
\bibliographystyle{utphys}
\bibliography{References}

\end{document}